\documentclass[12pt,showpacs,showkeys,amsmath,amssymb]{revtex4}
\usepackage{amsmath,amsfonts,amsthm,amscd,amssymb,latexsym}
\usepackage{bm}
\usepackage{dcolumn}
\usepackage{graphicx}
\usepackage{epstopdf}
\usepackage{color}
\usepackage{epsf}
\usepackage{epsfig}
\usepackage{graphicx, epic, eepic, color}
\usepackage{subfig}
\usepackage{braket}

\newcommand{\beq}{\begin{equation}}
\newcommand{\eeq}{\end{equation}}

\def\@{\partial_}
\def\be{\begin{equation}}
\def\ee{\end{equation}}

\def\negenspace{\kern-1.1em}

\def\sqr#1#2{{\vcenter{\hrule height.#2pt\hbox{\vrule width.#2pt
height#1pt \kern#1pt \vrule width.#2pt}\hrule height.#2pt}}}
\def\square{\mathchoice\sqr64\sqr64\sqr{4.2}3\sqr{3.0}3}

\begin{document}

\title{Local Limit of Nonlocal Gravity: A Teleparallel Extension of General Relativity}

\author{Javad \surname{Tabatabaei}$^{1}$}
\email{smj_tabatabaei@physics.sharif.edu}
\author{Shant \surname{Baghram}$^{1}$}
\email{baghram@sharif.edu}
\author{Bahram \surname{Mashhoon}$^{1,2,3}$} 
\email{mashhoonb@missouri.edu}

\affiliation{
$^1$Department of Physics, Sharif University of Technology, Tehran 11155-9161, Iran\\
$^2$School of Astronomy,
Institute for Research in Fundamental Sciences (IPM),
Tehran 19395-5531, Iran\\
$^3$Department of Physics and Astronomy,
University of Missouri, Columbia,
Missouri 65211, USA
}

\date{\today}

\begin{abstract}
We describe a general constitutive framework for a teleparallel extension of the general theory of relativity. This approach goes beyond the teleparallel equivalent of general relativity (TEGR) by broadening the analogy with the electrodynamics of media. In particular, the main purpose of this paper is to investigate in detail a local constitutive extension of TEGR that is the local limit of nonlocal gravity (NLG). Within this framework, we study the modified FLRW cosmological models. Of these, the most cogent turns out to be the modified Cartesian flat model which is shown to be inconsistent with the existence of a positive cosmological constant. Moreover, dynamic dark energy and other components of the modified Cartesian flat model evolve differently with the expansion of the universe as compared to the standard flat cosmological model.  The observational consequences of the modified Cartesian flat model are briefly explored and it is shown that the model is capable of resolving the $H_0$ tension.
\end{abstract}


\maketitle

\section{Introduction}

To explain current astronomical observations, dark matter is apparently necessary  to describe the dynamics of galaxies, clusters of galaxies, and structure formation in cosmology. Similarly, dark energy seems necessary to explain the accelerated expansion of the universe.  In the benchmark model of cosmology, the energy content of the universe comprises about 70\% dark energy, about 25\% dark matter and about 5\% visible matter. The physical nature of dark matter and dark energy is unknown at present. One or both dark aspects could conceivably be characteristic of the gravitational interaction. It therefore seems reasonable to attempt to modify Einstein's general relativity (GR) on the scales of galaxies and beyond in order to account for the observational data without any need for the dark content of the universe. To this end, many modified gravity theories have been proposed~\cite{Nojiri:2010wj, MGC, Nojiri:2017ncd}. This paper is about a certain constitutive extension of teleparallel gravity. The resulting theory can be described as the local limit of nonlocal gravity. 

Nonlocal gravity (NLG) is a classical nonlocal generalization of GR patterned after the nonlocal electrodynamics of media. To motivate this approach to the modification of GR, let us note that a postulate of locality runs through the special and general theories of relativity. In special relativity, Lorentz transformations are applied point by point along the world line of an accelerated observer in Minkowski spacetime in order to determine what the observer measures~\cite{Einstein}.  However, to measure the properties of electromagnetic waves, one must take their intrinsic nonlocal nature into account in accordance with the Huygens principle. Moreover, Bohr and Rosenfeld~\cite{BR} have pointed out that electromagnetic fields cannot be measured at a spacetime event; instead, a certain spacetime averaging procedure is necessary for this purpose. To extend the postulate of locality to the measurement of wave phenomena in Minkowski spacetime, one must consider the past history of the accelerated observer and this leads to nonlocal special relativity theory in which the observer's memory of its past acceleration is properly taken into account~\cite{NSR}. 

The locality postulate plays an essential role in Einstein's local principle of equivalence in rendering observers pointwise inertial in a gravitational field. The gravitational field equations in general relativity are thus partial differential equations. The intimate connection between inertia and gravitation, clearly revealed via Einstein's development of the general theory of relativity, implies that the universal gravitational interaction could be nonlocal as well. That is, the gravitational field equations could allow for the gravitational memory of past events in such a way that the local gravitational field would then satisfy partial integro-differential field equations.

Einstein's GR, as a field theory of gravitation, was modeled after Maxwell's electrodynamics. Maxwell's original field equations were formulated in a material medium in an inertial frame of reference. They involved the electromagnetic field strength given by $(\mathbf{E}, \mathbf{B}) \mapsto F_{\mu\nu}$ such that 
\begin{equation}\label{a}
F_{\mu \nu} = \partial_{ \mu}A_{ \nu} -\partial_{ \nu}A_{ \mu}\,,
\end{equation}
where $A_\mu$ is the vector potential. The response of the medium to the presence of $F_{\mu \nu}$ would be through its polarizability and magnetizability resulting in the net electromagnetic field excitations $(\mathbf{D}, \mathbf{H}) \mapsto H_{\mu\nu}$ such that 
\begin{equation}\label{b}
\partial_{\nu}H^{\mu \nu} = \frac{4 \pi}{c} J^{\mu}\,,
\end{equation}
where $J^\mu$ is the current 4-vector associated with \emph{free} electric charges in the inertial frame. In this theory, the electromagnetic fields  $F_{\mu \nu}$  and $H_{\mu\nu}$ are measured by the ideal inertial observers that are spatially at rest in the background inertial frame; moreover, the connection between these fields is through the \emph{constitutive relations} that are characteristic of the electromagnetic properties of the material  medium. For instance,   
$\mathbf{D} = \epsilon\, \mathbf{E}$ and $\mathbf{B} = \mu \,\mathbf{H}$ are the simplest forms of such relations, where $\epsilon$ and $\mu$ are the electric permittivity and magnetic permeability, respectively. However, even in Maxwell's time, experimental data indicated that the constitutive relations should be nonlocal~\cite{Hop}. Nonlocal electrodynamics of media is now an established part of standard physics; therefore, it appears natural to extend general relativity in a similar manner. That is, to introduce history dependence in gravitation, it is preferable in the first place to follow the approach that has been successful in electrodynamics. 

A nonlocal extension of general relativity theory has been developed  that is modeled after the nonlocal electrodynamics of media~\cite{Hehl:2008eu, Hehl:2009es}. The nonlocal constitutive kernel in the electrodynamics of media has its origin in atomic physics~\cite{L+L, Jack, HeOb}; however, no analogous atomic medium exists in the gravitational case. Therefore, the  nonlocal kernel in the gravitational case must be determined from observation. A comprehensive account of the resulting nonlocal gravity (NLG) theory is contained in Ref.~\cite{BMB}. A   significant observational consequence of this classical nonlocal generalization of Einstein's theory of gravitation is that the nonlocal aspect of gravity in the Newtonian regime of the theory appears to simulate dark matter~\cite{Rahvar:2014yta, Chicone:2015coa, Roshan:2021ljs, Roshan:2022zov, Roshan:2022ypk}.  

The classical nonlocal extension of GR can be accomplished through the framework of teleparallelism. Briefly, we start with GR and consider a gravitational field where events are characterized by an admissible system of spacetime coordinates $x^\mu$ with metric
\begin{equation}\label{I1}
ds^2 = g_{\mu \nu}\,dx^\mu \, dx^\nu\,.
\end{equation}
Free test particles and null rays follow geodesics in this spacetime manifold. Here, Greek indices run from 0 to 3, while Latin indices run from 1 to 3; moreover, the signature of the metric is +2. We use units such that $c = 1$, unless specified otherwise. 

In this spacetime, we choose a preferred set of observers with adapted tetrads $e^\mu{}_{\hat {\alpha}}(x)$ that are orthonormal; that is, 
\begin{equation}\label{I2}
 g_{\mu \nu}(x) \, e^\mu{}_{\hat {\alpha}}(x)\, e^\nu{}_{\hat {\beta}}(x)= \eta_{\hat {\alpha} \hat  {\beta}}\,.
\end{equation}
In our convention, hatted indices enumerate the tetrad axes in the local tangent space, while indices without hats are normal spacetime indices; moreover, $\eta_{\alpha \beta}$ is the Minkowski metric tensor given by diag$(-1,1,1,1)$.

We now employ our preferred tetrad frame field to define a nonsymmetric \emph{Weitzenb\"ock connection}~\cite{We}
\begin{equation}\label{I3}
\Gamma^\mu_{\alpha \beta}=e^\mu{}_{\hat{\rho}}~\partial_\alpha\,e_\beta{}^{\hat{\rho}}\,.
\end{equation}
This curvature-free connection is such that $\nabla_\nu\,e_\mu{}^{\hat{\alpha}}=0$, where $\nabla$ denotes covariant differentiation with respect to the Weitzenb\"ock connection. Our preferred tetrad frames are thus parallel throughout spacetime; that is, the  Weitzenb\"ock connection renders spacetime a 
parallelizable manifold. In this teleparallelism framework~\cite{Itin:2018dru, Maluf:2013gaa, Aldrovandi:2013wha}, distant vectors can be considered parallel if they have the same local components with respect to the preferred frame field. Furthermore,  the  Weitzenb\"ock connection is metric compatible, $\nabla_\nu\, g_{\alpha \beta}=0$, since the metric can be defined through the tetrad orthonormality relation. 

In our extended pseudo-Riemannian structure of GR, we have both the Levi-Civita and  Weitzenb\"ock connections that are compatible with the same spacetime metric $g_{\mu \nu}$. Curvature and torsion are basic tensors associated with a given connection. 
The symmetric  Levi-Civita connection is given by the Christoffel symbols
\begin{equation}\label{I4}
{^0}\Gamma^\mu_{\alpha \beta}= \frac{1}{2} g^{\mu \nu} (g_{\nu \alpha,\beta}+g_{\nu \beta,\alpha}-g_{\alpha \beta,\nu})\,.
\end{equation}
This  connection is torsion free, but has Riemannian curvature ${^0}R_{\alpha \beta \gamma \delta}$  that represents the gravitational field in GR.  In our convention, a left superscript ``0"  is used to refer to geometric quantities directly related to the Levi-Civita connection.  In particular, the gravitational field equations of GR can be expressed as~\cite{Einstein}
\begin{equation}\label{I5}
{^0}G_{\mu \nu} + \Lambda\, g_{\mu \nu}=\kappa\,T_{\mu \nu}\,, 
 \end{equation}
 where ${^0}G_{\mu \nu}$ is the Einstein tensor  
\begin{equation}\label{I6}
 {^0}G_{\mu \nu} := {^0}R_{\mu \nu}-\frac{1}{2} g_{\mu \nu}\,{^0}R\,.
 \end{equation} 
We denote the symmetric energy-momentum tensor of matter by $T_{\mu \nu}$; moreover,  $\Lambda$ is the cosmological constant and $\kappa:=8 \pi G/c^4$.  In Eq.~\eqref{I6}, the Ricci tensor
${^0}R_{\mu \nu} = {^0}R^{\alpha}{}_{\mu \alpha \nu}$ is the trace of the Riemann tensor and the scalar curvature ${^0}R = {^0}R^{\mu}{}_{\mu}$ is the trace of the Ricci tensor.

The spacetime \emph{torsion} tensor associated with the Weitzenb\"ock connection is given by
\begin{equation}\label{I7}
 C_{\mu \nu}{}^{\alpha}=\Gamma^{\alpha}_{\mu \nu}-\Gamma^{\alpha}_{\nu \mu}=e^\alpha{}_{\hat{\beta}}\Big(\partial_{\mu}e_{\nu}{}^{\hat{\beta}}-\partial_{\nu}e_{\mu}{}^{\hat{\beta}}\Big)\,.
\end{equation}
Furthermore, the difference between two connections on the same manifold is always a tensor. Hence, we define the \emph{contorsion} tensor
\begin{equation}\label{I8}
K_{\mu \nu}{}^\alpha= {^0} \Gamma^\alpha_{\mu \nu} - \Gamma^\alpha_{\mu \nu}\,.
\end{equation}
The metric compatibility of the Weitzenb\"ock connection means that contorsion is related to torsion; that is, 
\begin{equation}\label{I9}
K_{\mu \nu \rho} = \frac{1}{2}\, (C_{\mu \rho \nu}+C_{\nu \rho \mu}-C_{\mu \nu \rho})\,.
\end{equation}
The contorsion tensor is antisymmetric in its last two indices; in contrast, the torsion tensor is antisymmetric in its first two indices. It turns out that the curvature of the Levi-Civita connection  ${^0}R_{\mu \nu \rho \sigma}$ and the torsion of the Weitzenb\"ock connection $C_{\mu \nu \rho}$ are complementary aspects of the gravitational field in this extended framework~\cite{BMB}. 

It is important to point out a certain natural connection between the gravitational field strength $C_{\mu \nu \rho}$ and the electromagnetic field strength $F_{\mu \nu} = \partial_\mu A_\nu - \partial_\nu A_\mu$. Writing the torsion tensor as
\begin{equation}\label{I10}
 C_{\mu \nu}{}^{\hat{\alpha}}=e_\rho{}^{\hat{\alpha}}C_{\mu \nu}{}^{\rho}= \partial_{\mu}e_{\nu}{}^{\hat{\alpha}}-\partial_{\nu}e_{\mu}{}^{\hat{\alpha}}\,,
\end{equation}
we note that for each ${\hat{\alpha}}={\hat{0}}, {\hat{1}}, {\hat{2}}, {\hat{3}}$, we have here an analogue of the electromagnetic field tensor~\eqref{a} defined in terms of the vector potential $ A_\mu = e_{\mu}{}^{\hat{\alpha}}$. 

The traditional construction of the GR field equations is based on the Levi-Civita connection. On the other hand,  the Levi-Civita connection is the sum of the Weitzenb\"ock connection and the contorsion tensor in accordance with  Eq.~\eqref{I8}. Hence, GR field equations can be expressed in terms of the torsion tensor. This formulation of Einstein's theory is naturally analogous to Maxwell's electrodynamics. Indeed,  the result is the teleparallel equivalent of general relativity (TEGR) to which we now turn.

It is possible to express the Einstein tensor as~\cite{Hehl:2008eu, Hehl:2009es, BMB}
\begin{eqnarray}\label{G1}
 {^0}G_{\mu \nu}=\frac{\kappa}{\sqrt{-g}}\Big[e_\mu{}^{\hat{\gamma}}\,g_{\nu \alpha}\, \frac{\partial}{\partial x^\beta}\,\mathfrak{H}^{\alpha \beta}{}_{\hat{\gamma}}
-\Big(C_{\mu}{}^{\rho \sigma}\,\mathfrak{H}_{\nu \rho \sigma}
-\frac{1}{4}\,g_{\mu \nu}\,C^{\alpha \beta \gamma}\,\mathfrak{H}_{\alpha \beta \gamma}\Big) \Big]\,,
\end{eqnarray}
where the auxiliary torsion field $\mathfrak{H}_{\mu \nu \rho}$ is defined by 
\begin{equation}\label{G2}
\mathfrak{H}_{\mu \nu \rho}:= \frac{\sqrt{-g}}{\kappa}\,\mathfrak{C}_{\mu \nu \rho}\,, \qquad \mathfrak{C}_{\alpha \beta \gamma} :=C_\alpha\, g_{\beta \gamma} - C_\beta \,g_{\alpha \gamma}+K_{\gamma \alpha \beta}\,.
\end{equation}
The auxiliary torsion tensor $\mathfrak{C}_{\alpha \beta \gamma}$, where $C_\mu :=C^{\alpha}{}_{\mu \alpha} = - C_{\mu}{}^{\alpha}{}_{\alpha}$ is the torsion vector, is antisymmetric in its first two indices just like the torsion tensor. Moreover, as in GR, $g:=\det(g_{\mu \nu})$ and $\sqrt{-g}=\det(e_{\mu}{}^{\hat{\alpha}})$. Einstein's field equation~\eqref{I5} expressed in terms of torsion thus becomes the TEGR field equation 
\begin{equation}\label{G3}
 \frac{\partial}{\partial x^\nu}\,\mathfrak{H}^{\mu \nu}{}_{\hat{\alpha}}+\frac{\sqrt{-g}}{\kappa}\,\Lambda\,e^\mu{}_{\hat{\alpha}} =\sqrt{-g}\,(T_{\hat{\alpha}}{}^\mu + \mathbb{T}_{\hat{\alpha}}{}^\mu)\,,
\end{equation}
which is the analog of Maxwell's equation~\eqref{b}. Here,  $\mathbb{T}_{\mu \nu}$ is the traceless energy-momentum tensor of the gravitational field and is defined by
\begin{equation}\label{G4}
\sqrt{-g}\,\mathbb{T}_{\mu \nu} :=C_{\mu \rho \sigma}\, \mathfrak{H}_{\nu}{}^{\rho \sigma}-\frac 14  g_{\mu \nu}\,C_{\rho \sigma \delta}\,\mathfrak{H}^{\rho \sigma \delta}\,.
\end{equation}
The antisymmetry of $\mathfrak{H}^{\mu \nu}{}_{\hat{\alpha}}$ in its first two indices leads to the law of conservation of total energy-momentum tensor, namely, 
\begin{equation}\label{G5}
\frac{\partial}{\partial x^\mu}\,\Big[\sqrt{-g}\,(T_{\hat{\alpha}}{}^\mu + \mathbb{T}_{\hat{\alpha}}{}^\mu -\frac{\Lambda}{\kappa}\,e^\mu{}_{\hat{\alpha}})\Big]=0\,,
 \end{equation}
which follows from taking partial derivative $\partial/\partial x^\mu$ of Eq.~\eqref{G3}. Finally, it is important to note that while GR is based on the metric tensor $g_{\mu \nu}$,  TEGR is based on the orthonormal tetrad frame field $e^{\mu}{}_{\hat \alpha}(x)$ that is globally parallel via the Weitzenb\"ock connection. 

The teleparallel equivalent of general relativity (TEGR) is the gauge theory of the Abelian group of spacetime translations~\cite{Cho, BlHe, Puetzfeld:2019wwo}. Though nonlinear,  TEGR is therefore in a certain sense  analogous to Maxwell's equations in a medium with a simple constitutive relation~\cite{HeOb}. That is, $C_{\mu \nu}{}^{\hat{\alpha}}$ is, as noted before, similar to the electromagnetic field $F_{\mu \nu}$, where $(\mathbf{E},  \mathbf{B})\mapsto F_{\mu \nu}$, while,  $\mathfrak{H}_{\mu \nu}{}^{\hat \alpha}$  is similar to the electromagnetic excitation $H_{\mu \nu}$, where $(\mathbf{D},  \mathbf{H}) \mapsto H_{\mu \nu}$. Moreover, 
we can look upon Eq.~\eqref{G2}, namely,
\begin{equation}\label{G6}
\mathfrak{H}_{\alpha \beta \gamma} =  \frac{\sqrt{-g}}{\kappa}\,\mathfrak{C}_{\alpha \beta \gamma} = \frac{\sqrt{-g}}{\kappa}\,\left[\frac{1}{2}(C_{\gamma \beta \alpha } + C_{\alpha \beta \gamma} -C_{\gamma \alpha \beta}) +C_\alpha\, g_{\beta \gamma} - C_\beta \,g_{\alpha \gamma}\right]\,,
\end{equation} 
as the local constitutive relation of TEGR, since it connects $\mathfrak{H}_{\alpha \beta \gamma}$ to $C_{\alpha \beta \gamma}$. 

In describing  the electrodynamics of different media, the constitutive relations that relate the excitations $\mathbf{D}$ and $\mathbf{H}$ to the basic fields $\mathbf{E}$ and $\mathbf{B}$ are adjusted accordingly; however, Maxwell's basic field equations remain the same. We adopt this approach in extending TEGR; that is, the basic field equations of TEGR do not change, only the corresponding constitutive relation is appropriately modified. Therefore, we simply  change Eq.~\eqref{G6} in a suitable manner in the rest of this paper.

This approach differs from other extensions of  TEGR that involve, for instance,  the introduction of scalar fields into the theory.  That is,  scalar-torsion theories of gravity, which are analogues of scalar-tensor theories that extend GR, have been studied by a number of authors; for  recent reviews, see~\cite{Bahamonde:2021gfp, Hoh}. 

It is important to recognize that we could have arrived at TEGR using any other smooth orthonormal tetrad frame field $\lambda^{\mu}{}_{\hat \alpha}(x)$.  This circumstance is natural, since GR only depends on the metric tensor $g_{\mu \nu}$.  At each event with coordinates $x^\mu$, the two tetrad frame fields $\lambda^\mu{}_{\hat{\alpha}}(x)$ and $e^\mu{}_{\hat{\alpha}}(x)$  are related by a six-parameter  element of the \emph{local} Lorentz group  involving three boosts and three rotations; that is, 
$\lambda^\mu{}_{\hat{\alpha}}(x) = \mathcal{L}_{\hat \alpha}{}^{\hat \beta}(x)\, e^\mu{}_{\hat{\beta}}(x)$. This pointwise 6-fold degeneracy is generally removed when we modify the constitutive relation of TEGR. In our teleparallel extension of GR, the modified theory is then invariant only under the \emph{global} Lorentz group.


\section{Constitutive Extension of TEGR}

To maintain the analogy with  electrodynamics, we retain the gravitational field equations of TEGR, while the constitutive relation is modified. This means, in effect, that we replace $\mathfrak{H}_{\mu \nu \rho}$ in Eqs.~\eqref{G3} and~\eqref{G4} by $\mathcal{H}_{\mu \nu \rho}$ given by
\begin{equation}\label{T1}
\mathcal{H}_{\mu \nu \rho} = \frac{\sqrt{-g}}{\kappa}(\mathfrak{C}_{\mu \nu \rho}+ N_{\mu \nu \rho})\,,
\end{equation}
where $N_{\mu \nu \rho} = - N_{\nu \mu \rho}$  is a tensor that is related to the torsion tensor $C_{\mu \nu \rho}$.  For the moment, let us find the new extension of GR based on the new tensor field  $N_{\mu \nu \rho}$. 
The gravitational field equations take the form
\begin{equation}\label{T2}
 \frac{\partial}{\partial x^\nu}\,\mathcal{H}^{\mu \nu}{}_{\hat{\alpha}}+\frac{\sqrt{-g}}{\kappa}\,\Lambda\,e^\mu{}_{\hat{\alpha}} =\sqrt{-g}\,(T_{\hat{\alpha}}{}^\mu + \mathcal{T}_{\hat{\alpha}}{}^\mu)\,,
\end{equation}
where $\mathcal{T}_{\mu \nu}$ is the traceless energy-momentum tensor of the gravitational field in this case. We have
\begin{equation}\label{T3}
\kappa\,\mathcal{T}_{\mu \nu} = \kappa\,\mathbb{T}_{\mu \nu} + Q_{\mu \nu}\,,
\end{equation}
where $Q_{\mu \nu}$ is a traceless tensor  defined by
\begin{equation}\label{T4}
Q_{\mu \nu} := C_{\mu \rho \sigma} N_{\nu}{}^{\rho \sigma}-\frac 14\, g_{\mu \nu}\,C_{ \delta \rho \sigma}N^{\delta \rho \sigma}\,.
\end{equation} 
The total energy-momentum conservation law can now be expressed as
\begin{equation}\label{T5}
\frac{\partial}{\partial x^\mu}\,\Big[\sqrt{-g}\,(T_{\hat{\alpha}}{}^\mu + \mathcal{T}_{\hat{\alpha}}{}^\mu -\frac{\Lambda}{\kappa}\,e^\mu{}_{\hat{\alpha}})\Big]=0\,.
 \end{equation}
 
 To find the modified GR field equations, let us start with Eq.~\eqref{T1} and substitute
 \begin{equation}\label{T6}
\mathfrak{H}_{\mu \nu \rho} = \mathcal{H}_{\mu \nu \rho} - \frac{\sqrt{-g}}{\kappa} N_{\mu \nu \rho}\,
\end{equation}
 in the Einstein tensor~\eqref{G1} to get 
 \begin{equation}\label{T7}
^{0}G_{\mu \nu} + \Lambda g_{\mu \nu} = \kappa T_{\mu \nu}   + \mathbb{R}_{\mu \nu}\,, \qquad   \mathbb{R}_{\mu \nu} := Q_{\mu \nu} -  \mathcal{N}_{\mu \nu}\,,
\end{equation}
where  we have used Eq.~\eqref{T2}. Here, $\mathcal{N}_{\mu \nu}$ is a  tensor defined by
\begin{equation}\label{T8}
\mathcal{N}_{\mu \nu} = g_{\nu \alpha} e_\mu{}^{\hat{\gamma}} \frac{1}{\sqrt{-g}} \frac{\partial}{\partial x^\beta}\,(\sqrt{-g}N^{\alpha \beta}{}_{\hat{\gamma}})\,.
\end{equation} 

It is natural to split the modified GR field equations into its symmetric and antisymmetric parts; that is, we have the modified Einstein equations 
\begin{equation}\label{T9}
^{0}G_{\mu \nu} +   \Lambda g_{\mu \nu}  = \kappa T_{\mu \nu} +  \mathbb{R}_{(\mu \nu)}\,, \qquad \mathbb{R}_{(\mu \nu)} = Q_{(\mu \nu)}  - \mathcal{N}_{(\mu \nu)}\,
\end{equation}
and the constraint equations
\begin{equation}\label{T10}
\mathbb{R}_{[\mu \nu]} = 0 \,, \qquad Q_{[\mu \nu]} = \mathcal{N}_{[\mu \nu]}\,.
\end{equation}
We thus have 16 field equations for the 16 components of the tetrad frame field. Of the 16 components of the fundamental tetrad $e^\mu{}_{\hat{\alpha}}$,  10 fix the components of the metric tensor $g_{\mu \nu}$ via the orthonormality condition, while the other 6 are local Lorentz degrees of freedom (i.e., boosts and rotations). Similarly,  the 16 field equations of modified GR for the 16 components of the fundamental tetrad $e^\mu{}_{\hat{\alpha}}$ naturally split into the 10 modified Einstein equations  plus the 6 integral constraint equations for the new tensor $N_{\mu \nu \rho}$.

We have here a general constitutive framework for the teleparallel extension of GR.  It remains to specify the exact connection between $N_{\mu \nu \rho}$ and $C_{\mu \nu \rho}$. A nonlocal relation has led to nonlocal gravity (NLG) theory~\cite{Hehl:2008eu, Hehl:2009es, BMB}.   The local limit of this nonlocal relation is the main focus of the present investigation.


\subsection{Nonlocal Gravity}

In the physical description of the electrodynamics of a medium at rest in an inertial frame of reference, Maxwell's original field equations hold together with constitutive relations that connect the components of $H_{\mu \nu}$ and $F_{\mu \nu}$ as measured by fundamental inertial observers that remain at rest with the medium. Similarly,  in nonlocal gravity (NLG), we assume, in close analogy with the nonlocal electrodynamics of media, that the components of $N_{\mu \nu \rho}$, as measured by the fundamental observers of the theory with adapted tetrads $e^\mu{}_{\hat{\alpha}}$, must be physically related to the corresponding measured components of $X_{\mu \nu \rho}$ that is directly connected to the torsion tensor~\cite{BMB, Puetzfeld:2019wwo, Mashhoon:2022ynk}. That is,
\begin{equation}\label{N1}
N_{\hat \mu \hat \nu \hat \rho}(x) =  \int  \mathcal{K}(x, x')\,X_{\hat \mu  \hat \nu  \hat \rho }(x') \sqrt{-g(x')}\, d^4x' \,,
\end{equation} 
where $\mathcal{K}(x, x')$ is the basic causal kernel of NLG and
\begin{equation}\label{N2}
X_{\hat \mu \hat \nu \hat \rho}= \mathfrak{C}_{\hat \mu \hat \nu \hat \rho}+ \check{p}\,(\check{C}_{\hat \mu}\, \eta_{\hat \nu \hat \rho}-\check{C}_{\hat \nu}\, \eta_{\hat \mu \hat \rho})\,.
\end{equation}
Here, $\check{p}\ne 0$ is a constant dimensionless parameter and  $\check{C}^\mu$ is the torsion pseudovector defined via the Levi-Civita tensor $\epsilon_{\alpha \beta \gamma \delta}$ by
\begin{equation}\label{N3}
\check{C}_\mu :=\frac{1}{3!} C^{\alpha \beta \gamma}\,\epsilon_{\alpha \beta \gamma \mu}\,.
\end{equation}

In connection with the parity violating term proportional to $\check{p}$, it has been shown in detail in~\cite{BMB} that the field equations of \emph{linearized} NLG are mathematically inconsistent in the absence of this term. The analysis of linearized field equations of NLG reveals that the parity violating term may only contribute in highly time-dependent situations. Further investigation is necessary to clarify the status of parity violation within the full nonlinear regime of NLG.  Beyond linearized NLG, no exact nonlinear solution of nonlocal gravity is known at present; in fact, some of the difficulties have been discussed in~\cite{Bini:2016phe}.  Indeed, parity-violating solutions may exist in which the torsion pseudovector $\check{C}_\mu$ that appears in the constitutive relation of NLG may be nonzero. Theories that exhibit gravitational parity violation have been of current interest~\cite{Nishizawa:2018srh, Conroy:2019ibo}. 

It is important to remark that the only known exact solution of NLG is the trivial solution, namely, we recover Minkowski spacetime in the absence of the gravitational field.  That is, with $e^{\mu}{}_{\hat \alpha} = \delta^\mu_{\alpha}$, $g_{\mu \nu} = \eta_{\mu \nu}$, $T_{\mu \nu} = 0$ and $\Lambda = 0$ we have an exact solution of NLG. Thus far, it has only been possible to show that de Sitter solution is not an exact solution of NLG~\cite{Mashhoon:2022ynk}. Linearized NLG has been treated in~\cite{BMB}; moreover, nonlocal Newtonian cosmology has been treated within the Newtonian regime of NLG~\cite{Chicone:2015sda, Chicone:2017oqt}.  

There are many other nonlocal models of gravity and cosmology; see, for instance~\cite{Woodard:2018gfj, Deser:2019lmm, Balakin:2022gjw, BasiBeneito:2022wux, Jusufi:2023ayv} and the references cited therein. 


\subsection{Local Limit of NLG}

The local limit of Eq.~\eqref{N1} can be obtained by assuming that the kernel is proportional to the 4D Dirac delta function, namely,  
\begin{equation}\label{N4}
\mathcal{K}(x, x') := \frac{S(x)}{\sqrt{-g(x)}}\,\delta(x-x')\,,
\end{equation}
where $S(x)$ is a dimensionless scalar function that must be determined on the basis of observational data.  
In this case, the nonlocal constitutive relation~\eqref{N1} reduces to  
\begin{equation}\label{N5}
N_{\mu \nu \rho}(x)  = S(x) X_{\mu \nu \rho} = S(x)\,[\mathfrak{C}_{\mu \nu \rho}(x) + \check{p}\,(\check{C}_\mu\, g_{\nu \rho}-\check{C}_\nu\, g_{\mu \rho})]\,.
\end{equation} 
Moreover, Eq.~\eqref{T1} takes the form
\begin{equation}\label{N6}
\mathcal{H}_{\mu \nu \rho} = \frac{\sqrt{-g}}{\kappa}[(1+S)\,\mathfrak{C}_{\mu \nu \rho}+ S\,\check{p}\,(\check{C}_\mu\, g_{\nu \rho}-\check{C}_\nu\, g_{\mu \rho})]\,.
\end{equation}
If $S(x) = 0$, we recover TEGR; otherwise, we have a generalization of GR. Of course,  the local and linear relation~\eqref{N5} can be generalized; that is, 
\begin{equation}\label{N7}
N_{\mu \nu \rho}(x)  = \frac{1}{2} \chi_{\mu \nu \rho}{}^{\alpha \beta \gamma}(x) X_{\alpha \beta \gamma}(x)\,.
\end{equation}
Such  six-index gravitational constitutive tensors as $\chi_{\mu \nu \rho}{}^{\alpha \beta \gamma}$ have been studied and classified in~\cite{Itin:2018dru}. 

The new local constitutive relation enlarges TEGR, which is equivalent to a pure spin-2 theory, namely, GR, by the addition of a scalar function. It is clear from Eq.~\eqref{N6} that $1+S > 0$; otherwise, the new theory will not have GR as a limit. There is no field equation for  $S(x)$; however, we may be able to select $S(x)$ on the basis of certain consistency conditions in analogy with the electrodynamics of media. Eventually, we have to  determine
 $S(x)$ from observation, just as in NLG we must ultimately determine the fundamental nonlocal kernel of the theory from the comparison of the theory with observational data~\cite{Rahvar:2014yta, Chicone:2015coa, Roshan:2021ljs, Roshan:2022zov}. In the local limit of NLG, we have the prospect of finding exact solutions that explore strong field regimes involving cosmological models and black holes.


\subsection{Analogy with the Electrodynamics of Media}

In the electrodynamics of media, especially magnetic media, the phenomena associated with hysteresis cannot be ignored; therefore, the constitutive relations are in general nonlocal~\cite{Jack}. However, in most applications of the electrodynamics of media, one uses the simple relations $\mathbf{D} = \epsilon(x) \mathbf{E}$ and $\mathbf{B} = \mu (x) \mathbf{H}$. Presumably, these local limits of nonlocal constitutive relations of linear media capture some important aspects of the general nonlocal problem. The same might be expected to hold in the gravitational case. We could partially compensate for the lack of exact solutions of NLG in the case of strong gravitational fields by searching for solutions of the new local theory in the areas of cosmology and black hole physics. 

The local quantities $\epsilon(x)$ and $\mu (x)$ are characteristics of the medium in electrodynamics; similarly, $S(x)$ is characteristic of the background spacetime in the new local theory. The functional form of $S(x)$ must therefore be consistent with the nature of the background spacetime. In analogy with electrodynamics, we call $S(x)$ the \emph{susceptibility} function and tentatively call the new theory ``Modified TEGR", since we simply add an extra term to the constitutive relation of TEGR.  Just as electric permittivity and magnetic permeability can pick out an electromagnetic medium, $S(x)$ can characterize a spacetime in modified TEGR. For instance, a spatially homogeneous time dependent spacetime in our modified TEGR is endowed with an independent function $S(t)$ such that $dS/dt \ne 0$. 

It must be mentioned that other modified teleparallel theories of gravity have been considered by a number of authors; 
see, for instance~\cite{Ferraro:2006jd, Maluf:2011kf, Bahamonde:2015zma, Boehmer:2021aji,  Capozziello:2022zzh} and the references cited therein. 

At present, it is not known whether NLG or its local limit (namely, modified TEGR) can be derived from an action principle. Indeed, many important physical systems have dynamics that cannot be derived from stationary action principles. Dissipative processes generally lack Lagrangian descriptions. An example of this situation in fluid dynamics is the Navier-Stokes system~\cite{Bekenstein:2014uwa}. 
In the electrodynamics of media, averaging procedures are necessary to determine the contributions of the polarization and magnetization of the medium to the electromagnetic field. As a consequence, there is no classical field theory Lagrangian in the literature for the electrodynamics of media. This circumstance extends to \emph{nonlocal} electrodynamics of media as well.
In the gravitational case, the field equations of nonlocal gravity (NLG) include a certain average of the gravitational field over past events; in essence, it is this nonlocal aspect of the theory that simulates effective dark matter. The analogy with the electrodynamics of media suggests that theories that modify general relativity in this way do not have Lagrangian descriptions. Previous work in this direction implies that an action principle for NLG would be in conflict with causality~\cite{Hehl:2009es}. In the local limit of nonlocal gravity considered in the present work, the averaging is replaced by a local weight function $S(x)$ that is  reminiscent of $\epsilon (x)$ and $\mu(x)$ of the local electrodynamics of media. Can the field equations for such a modified theory of gravitation be derived from an action principle? This seems rather unlikely based on the analogy with the electrodynamics of media, but a definitive answer is not known at present. 

\subsection{Modified TEGR}

To gain insight into the nature of modified TEGR, we must solve its field equations which are obtained from the substitution of the auxiliary torsion field~\eqref{N6} in  Eq.~\eqref{T2}. This is a tetrad theory; therefore, the solution of the field equations would consist of the 16 components  of an orthonormal tetrad frame field $e^\mu{}_{\hat{\alpha}}(x)$ adapted to the preferred observers in spacetime. As in NLG, we recover Minkowski spacetime in the absence of gravity. That is, the tetrad frame field $e^\mu{}_{\hat{\alpha}}(x) = \delta^{\mu}_{\alpha}$ adapted to preferred static inertial observers with Cartesian coordinates $x^\mu$ in Minkowski spacetime is an exact solution of modified TEGR with $T_{\mu \nu} = 0$ and $\Lambda=0$, regardless of $S(x)$.

\section{Modified TEGR: Linear Approximation}

 In this section, we look for an approximate solution of modified TEGR field equations with $S(x) \ne 0$ and 
 $\Lambda=0$ that is a first-order perturbation about Minkowski spacetime. The purpose of this section is to develop and study the resulting general linear weak-field solution of modified TEGR.

\subsection{Linearization}

We begin by linearizing the theory about Minkowski spacetime; that is, we assume the fundamental frame field of the theory is given by~\cite{BMB, Mashhoon:2019jkq} 
\begin{equation}\label{L1}
 e_\mu{}^{\hat{\alpha}}=\delta_\mu ^{\alpha}+\psi^{\alpha}{}_\mu\,, \qquad  e^\mu{}_{\hat{\alpha}}=\delta^\mu _{\alpha} -\psi^\mu{}_{\alpha}\,.
\end{equation}
In the linear perturbing field $\psi_{\mu \nu}(x)$, the distinction between spacetime and tetrad indices can be ignored at this level of approximation. Here, the 16 components of $\psi_{\mu \nu}$ represent the gravitational potentials of a finite source of mass-energy that is \emph{at rest} in a compact region of space. We thus break the invariance of the theory under the  global Lorentz group by fixing the background inertial frame of reference to be the rest frame of the source. We define the symmetric and antisymmetric components of $\psi_{\mu \nu}$ by
\begin{equation}\label{L2}
 h_{\mu \nu}:=2\psi_{(\mu \nu)}\,, \qquad  \phi_{\mu \nu}:=2\psi_{[\mu \nu]}\,.
\end{equation}
The tetrad orthonormality condition then implies
\begin{equation}\label{L3} 
g_{\mu \nu}=\eta_{\mu \nu} + h_{\mu \nu}\,.
\end{equation}
As in GR, we introduce the  the trace-reversed potentials
\begin{equation}\label{L4}
\bar{h}_{\mu \nu}=h_{\mu \nu}-\frac{1}{2}\,\eta_{\mu \nu}h\,, \qquad   h:=\eta_{\mu \nu}h^{\mu \nu}\,,
\end{equation} 
where $\bar{h}=-h$ and 
\begin{equation}\label{L5}
\psi_{\mu \nu}=\frac{1}{2}\,\bar{h}_{\mu \nu}+\frac{1}{2}\,\phi_{\mu \nu}-\frac{1}{4}\,\eta_{\mu \nu}\,\bar{h}\,.
\end{equation} 

The gravitational potentials are now in suitable form to calculate the gravitational field quantities to first order in the perturbation. The linearized torsion tensor is 
\begin{equation}\label{L6}
C_{\mu \nu \sigma}=\partial_\mu \psi_{\sigma \nu}-\partial_\nu \psi_{\sigma \mu}
\end{equation}
and the torsion vector and pseudovector are given by
\begin{equation}\label{L7}
C_{\mu}=\frac{1}{4}\, \partial_\mu \bar{h} + \frac{1}{2}\, \partial_\nu(\bar{h}^{\nu}{}_\mu+\phi^{\nu}{}_\mu)\,, \qquad  \check{C}^\mu = \frac{1}{6}\, \epsilon^{\mu \nu \rho \sigma}\,\phi_{\nu \rho, \sigma}\,.
\end{equation}
Similarly, the auxiliary torsion tensor is
\begin{equation}\label{L8}
\mathfrak{C}_{\mu \sigma \nu}=-\bar{h}_{\nu [\mu,\sigma]}-\eta_{\nu [\mu}\bar{h}_{\sigma ]\rho,}{}^\rho+\frac{1}{2}\,\phi_{\mu \sigma, \nu}+\eta_{\nu [\mu} \phi_{\sigma ] \rho,}{}^\rho\,,
\end{equation}
and the Einstein tensor can be written as 
\begin{equation}\label{L9}
^{0}G_{\mu \nu}=\partial_\sigma \mathfrak{C}_{\mu}{}^{\sigma}{}_{\nu}=
-\frac{1}{2}\,\Box \,\bar{h}_{\mu \nu}+\bar{h}^\rho{}_{(\mu,\nu)\rho}
-\frac{1}{2}\,\eta_{\mu \nu}\bar{h}^{\rho \sigma}{}_{,\rho \sigma}\,,
\end{equation}
where $\Box :=\eta^{\alpha \beta}\partial_\alpha \partial_\beta$ and  $\partial_{\nu}\, ^{0}G^{\mu \nu}=0$, since the auxiliary torsion tensor is antisymmetric in its first two indices.

Let us recall that in general the linearized form of modified GR field equations~\eqref{T9} and~\eqref{T10} are given by~\cite{BMB}
\begin{equation}\label{L10}
 ^{0}G_{\mu \nu}+\frac{1}{2}\, \partial_\sigma\,(N_{\mu}{}^{\sigma}{}_{\nu}+N_{\nu}{}^{\sigma}{}_{\mu})= \kappa\,  T_{\mu \nu}\,              
\end{equation}
and
\begin{equation}\label{L11}
  \partial_\sigma\,N_{\mu}{}^{\sigma}{}_{\nu}= \partial_\sigma\, N_{\nu}{}^{\sigma}{}_{\mu}\,,              
\end{equation}
respectively. These imply, just as in GR, the energy-momentum conservation law for mass-energy, namely,  $\partial_{\nu} T^{\mu \nu}=0$. Writing  Eq.~\eqref{N5} as
\begin{equation}\label{L12}
N_{\mu}{}^{\sigma}{}_{\nu}= S(x) X_{\mu}{}^{\sigma}{}_{\nu}(x)\,,                 
\end{equation}
the modified GR field equations become
\begin{equation}\label{L13}
 ^{0}G_{\mu \nu}(x) + \partial_\sigma\,[S(x)\,X_{(\mu}{}^{\sigma}{}_{\nu)}(x)] = \kappa\,  T_{\mu \nu}(x)\,              
\end{equation}
and
\begin{equation}\label{L14}
 \partial_\sigma\,[S(x)\, X_{[\mu}{}^{\sigma}{}_{\nu]}(x)] =0\,.
\end{equation}
With $X_{\mu \sigma \nu}$ given by Eq.~\eqref{N2} and $\check{p} \ne 0$, we find that in the \emph{linear} regime,
\begin{equation}\label{L15}
X_{(\mu}{}^{\sigma}{}_{\nu)} = \mathfrak{C}_{(\mu}{}^{\sigma}{}_{\nu)}+ \check{p}\,\big[\check{C}_{(\mu}\delta^\sigma_{\nu)}-\check{C}^\sigma \eta_{\mu \nu}\big]\,, \qquad X_{[\mu}{}^{\sigma}{}_{\nu]}= 
\mathfrak{C}_{[\mu}{}^{\sigma}{}_{\nu]}+ \check{p}\,\check{C}_{[\mu}\delta^\sigma_{\nu]}\,.
\end{equation}
The torsion pseudovector $\check{C}^\sigma$ is the dual of $C_{[\mu \nu \rho]}$; moreover, in our linear approximation scheme  $C_{[\mu \nu \rho]}=-\phi_{[\mu \nu , \rho]}$ and $\check{C}^{\sigma}{}_{,\sigma}=0$. Thus the part of the constitutive relation proportional to $\check{p}$ is given exclusively by the derivatives of  the antisymmetric tetrad potentials $\phi_{\mu \nu}$ and vanishes for $\phi_{\mu \nu}=0$. Using $^{0}G_{\mu \nu}=\partial_\sigma \mathfrak{C}_{(\mu}{}^{\sigma}{}_{\nu)}$ and $\partial_\sigma \mathfrak{C}_{[\mu}{}^{\sigma}{}_{\nu]} = 0$, the field equations take the form
\begin{equation}\label{L16}
(1+S)\, ^{0}G_{\mu \nu}(x) + S_{,\sigma}\,X_{(\mu}{}^{\sigma}{}_{\nu)}(x)+ S \check{p}\,\check{C}_{(\mu , \nu)} = \kappa\,  T_{\mu \nu}(x)\,              
\end{equation}
and
\begin{equation}\label{L17}
 S_{,\sigma}\,X_{[\mu}{}^{\sigma}{}_{\nu]}(x)+ S \check{p}\,\check{C}_{[\mu , \nu]} =0\,.
\end{equation}
To have regular second-order field equations, we must have $1+S(x) > 0$. 

Let us briefly digress here and consider the possibility that $1+S = 0$. It is simple to see that source-free (i.e., $T_{\mu \nu} = 0$ and $\Lambda = 0$) linearized field equations~\eqref{L16} and~\eqref{L17} are satisfied  for $\phi_{\mu \nu} = 0$ and $S = -1$. Extending this result to the nonlinear case is not so simple; that is, the general source-free field equations are satisfied for $S = -1$ provided 
$\check{C}_\mu = 0$ and $Q_{\mu \nu} = 0$. It is very possible that no such solution exists. The general field equations of modified TEGR for $S = -1$ are first-order partial differential equations that have a peculiar form. It seems that there is no reasonable spacetime that could satisfy these conditions. 

The gravitational potentials $\psi_{\mu \nu}$ are gauge dependent. Under an infinitesimal coordinate transformation, $x^\mu \mapsto x'^\mu=x^\mu-\epsilon^\mu(x)$, we find to linear order in $\epsilon^\mu$,  $\psi_{\mu \nu} \mapsto \psi'_{\mu \nu}=\psi_{\mu \nu}+\epsilon_{\mu,\nu}$. Therefore,
\begin{equation}\label{L18}
\bar{h}'_{\mu \nu}=\bar{h}_{\mu \nu}+\epsilon_{\mu,\nu}+\epsilon_{\nu,\mu}-\eta_{\mu \nu}\epsilon^\alpha{}_{,\alpha}\,, \qquad   \phi'_{\mu \nu}=\phi_{\mu \nu}+\epsilon_{\mu,\nu}-\epsilon_{\nu,\mu}
\end{equation} 
and $\bar{h}'=\bar{h}-2\epsilon^\alpha{}_{,\alpha}$. On the other hand,  the gravitational field tensors $C_{\mu \nu \rho}$ and $\mathfrak{C}_{\mu\nu \rho}$ as well as the gravitational field equations are gauge invariant, as expected. 

In general, we can impose the transverse gauge condition
\begin{equation}\label{L19}
\bar{h}^{\mu\nu}{}_{, \nu}=0\,,
\end{equation}
which does not completely fix the gauge. We can still use four functions $\epsilon^\mu$ such that $\square\,\epsilon^\mu = 0$. With the transverse gauge condition, we have
\begin{equation}\label{L20}
^{0}G_{\mu \nu} = -\frac{1}{2}\,\square\,\bar{h}_{\mu\nu}\,,
\end{equation}
which simplifies Eq.~\eqref{L16}. 
 
\subsection{ Newtonian Limit}

In this limit, we consider weak fields and slow motions; moreover, we can formally let $c \to \infty$. As in Ref.~\cite{Mashhoon:2019jkq},  we  assume the transverse gauge condition holds and  $\phi_{\mu \nu} = 0$. Furthermore, we assume $S(x)$ is a constant given by $\mathbb{S}_{\rm N}$. A detailed examination of the field Eqs.~\eqref{L16} and~\eqref{L17} reveals that only the $ \mu = \nu = 0$ case is important with $\bar{h}_{0 0} = -4 \Phi/c^2$ and $T_{00} = \rho\, c^2$, where $\Phi$ is the Newtonian gravitational potential and $\rho$ is the density of matter~\cite{BMB, Mashhoon:2019jkq, Bini:2021gdb}. 
The background is the static Newtonian space and time, a circumstance that is in conformity with the constancy of the susceptibility $\mathbb{S}_{\rm N}$. We find from Eq.~\eqref{L16} that
\begin{equation}\label{P1}
\nabla^2\,\Phi (x) = \frac{4 \pi G \rho}{1+\mathbb{S}_{\rm N}}\,
\end{equation}
and the second field equation disappears. In this Newtonian framework, the force of gravity on a particle of mass $m$ is given by $\mathbf{F} = - m\,\nabla \Phi$. Thus, the gravitational force between two point masses has the Newtonian form except that it is augmented by a constant factor, namely, $(1+\mathbb{S}_{\rm N})^{-1} > 0$. 

To interpret Eq.~\eqref{P1} in terms of effective dark matter, let us  assume that 
\begin{equation}\label{P2}
(1+\mathbb{S}_{\rm N}) (1+\mathbb{Q}_{\rm N})  = 1\,, \qquad \mathbb{S}_{\rm N}\in (-1, 0]\,, \qquad \mathbb{Q}_{\rm N} \in [0, \infty)\,.
\end{equation}
Then, we can write the modified Poisson Eq.~\eqref{P1} as
\begin{equation}\label{P3}
\nabla^2\,\Phi (x) = 4 \pi G (\rho + \rho_D)\,, \qquad \rho_D = \mathbb{Q}_{\rm N}\, \rho\,,
\end{equation}
where $\rho_D$ has the interpretation of the density of dark matter in this framework. For a point mass $m$, say,  the dark matter associated with $m$ is located at the point mass and has magnitude $\mathbb{Q}_{\rm N} \,m$. In the Newtonian regime of nonlocal gravity (NLG), the density of dark matter is the convolution of the density of matter with a spherically symmetric reciprocal kernel. In the local limit of NLG in the Newtonian regime, the folding with the reciprocal kernel reduces to multiplication by a constant $\mathbb{Q}_{\rm N}$. That is, for a point mass, the static spherical cocoon of effective dark matter associated with a point mass in NLG~\cite{Roshan:2021ljs, Roshan:2022zov} collapses on the point particle in the local limit of NLG. Therefore, in contrast to the Newtonian regime of NLG~\cite{Rahvar:2014yta, Roshan:2022zov}, the local limit of NLG is not capable of explaining the rotation curves of spiral galaxies. 

We therefore abandon Eqs.~\eqref{P2}--\eqref{P3} and consider an alternative interpretation of Eq.~\eqref{P1} in terms of Newtonian gravity but with the gravitational constant $G \to (1+\mathbb{S}_{\rm N})^{-1}\,G$. The Newtonian constant of gravitation has been measured with a relative standard uncertainty of about 20 parts per million~\cite{Luo:2020gjh}. Hence, to maintain consistency with the physics of the solar system, we must have $|\mathbb{S}_{\rm N}| < 2 \times 10^{-5}$. 

\subsection{Free Gravitational Waves}

In linearized nonlocal gravity, free gravitational waves satisfy the nonlocal gravitational wave equation~\cite{BMB}
\begin{equation}\label{W1}
\square\,\bar{h}_{ij}(x) + \int W(x-y) \, \bar{h}_{ij,0}(y)\,d^4y = 0\,,
\end{equation}
where $W$ is a certain kernel of NLG. This result follows from the source-free linearized field equations once the gauge conditions $\bar{h}^{\mu\nu}{}_{, \nu}=0$, $\bar{h}_{0\mu} = 0$ and $\phi_{\mu \nu} = 0$ are imposed.  The nonlocal wave Eq.~\eqref{W1} is reminiscent of a damped oscillator whose velocity would be represented by  $\partial \bar{h}_{ij}/\partial t$. The wave amplitude decays as the wave propagates due to fading memory in NLG. 

In the case of source-free field Eqs.~\eqref{L16} and~\eqref{L17}, we can simply impose the transverse gauge condition and $\phi_{\mu \nu} = 0$. To go further in analogy with Eq.~\eqref{W1}, we suppose that $S$ is only a function of time, i.e. $S = S(t)$.  With these assumptions, the field equations then become
\begin{equation}\label{W2}
 (1+S)\square\,\bar{h}_{\mu\nu} + \frac{dS}{dt} [\bar{h}_{\mu \nu,}{}^{0} - \bar{h}_{(\nu}{}^{0}{}_{, \mu)}] = 0\,,
\end{equation}
\begin{equation}\label{W3}
 \frac{dS}{dt} \bar{h}_{[\mu}{}^{0}{}_{,\nu]} = 0\,.
\end{equation}
It is now possible to enforce the additional gauge conditions $\bar{h}_{0\mu} = 0$. In this case, of Eqs.~\eqref{W2} and~\eqref{W3} only the former remains in the form of a wave equation for  $\bar{h}_{ij}$.   Here, we have used 
\begin{equation}\label{W4}
\mathfrak{C}_{\mu}{}^{\sigma}{}_{\nu} = -\frac{1}{2} \bar{h}_{\mu \nu ,}{}^\sigma  + \frac{1}{2} \bar{h}_{\nu}{}^{\sigma}{}_{, \mu}\,
\end{equation}
and 
\begin{equation}\label{W5}
\mathfrak{C}_{(\mu}{}^{\sigma}{}_{\nu)} = -\frac{1}{2} \bar{h}_{\mu \nu ,}{}^\sigma  + \frac{1}{2} \bar{h}_{(\nu}{}^{\sigma}{}_{, \mu)}\,, \qquad  \mathfrak{C}_{[\mu}{}^{\sigma}{}_{\nu]} = - \frac{1}{2}\bar{h}_{[\mu}{}^{\sigma}{}_{,\nu]}\,.
\end{equation}

 For $\bar{h}_{ij}$, Eq.~\eqref{W2} implies
\begin{equation}\label{W6}
 \square\,\bar{h}_{ij} - \bar{ \gamma}(t) \frac{\partial \bar{h}_{ij}}{\partial t} = 0\,,
\end{equation}
where 
\begin{equation}\label{W7}
\bar{\gamma}(t) =  \frac{1}{1+S}\frac{dS}{dt} = \frac{d}{dt}\ln(1+S)\,.
\end{equation}
The transverse gauge condition reduces to $\bar{h}^{ij}{}_{, j}=0$, which is consistent with the propagation Eq.~\eqref{W6}. 

To solve Eq.~\eqref{W6}, we assume that each component of  the wave function $\bar{h}_{ij}$ has the form
\begin{equation}\label{W8}
 \mathbb{H}(t) e^{i \bar{\mathbf{k}}\cdot \mathbf{x}}\,,
\end{equation}
where $\mathbb{H}$ satisfies a harmonic oscillator equation with time-dependent damping. That is, 
\begin{equation}\label{W9}
 \frac{d^2\mathbb{H}}{dt^2} + \bar{k}^2 \mathbb{H} + \bar{\gamma}(t) \frac{d\mathbb{H}}{dt} = 0\,,
\end{equation}
where $\bar{k} = |\bar{\mathbf{k}}|$. For a positive constant $\bar{\gamma}$, we have the standard damped harmonic oscillator. On the other hand, the solution will always be damped if $\bar{\gamma}(t) > 0$, since
the energy associated with the harmonic oscillator constantly decays, namely,  
\begin{equation}\label{W10}
\frac{d}{dt}\left[\frac{1}{2} \left(\frac{d\mathbb{H}}{dt}\right)^2 + \frac{1}{2} \bar{k}^2 \mathbb{H}^2\right] = - \bar{\gamma}(t) \left(\frac{d\mathbb{H}}{dt}\right)^2\,.
\end{equation}

As in NLG~\cite{BMB, Chicone:2012cc, Mashhoon:2013jqa}, free gravitational waves in our local limit of NLG are indeed damped provided $dS/dt > 0$, since $1+S > 0$.  A comment is in order here regarding the fact that in a dynamic time-dependent spacetime, $dS/dt > 0$  appears to be a natural requirement in order to maintain $1+S > 0$; that is, $dS/dt > 0$ ensures that an initially positive $1+S$ will remain positive for all time. The Wentzel-Kramers-Brillouin (WKB) treatment of wave Eq.~\eqref{W6} is contained in Appendix A. 

The rest of this paper is devoted to finding the simplest exact cosmological models of modified TEGR. The current benchmark model of cosmology is the flat FLRW solution; therefore, we focus on the way the FLRW model is modified in our approach. We begin with the simplest modified conformally flat spacetimes. 

\section{Modified TEGR: Conformally Flat Spacetimes}

We are interested in exact solutions of modified TEGR that have a conformally flat metric
\begin{equation}\label{M1}
ds^2= e^{2U} \, \eta_{\mu \nu}\, dx^\mu\,dx^\nu\,,
\end{equation}
where $x^\mu = (\eta, x^i)$ and $U(x)$ is a scalar. We use \emph{conformal time} $x^0 = \eta$ in this section to agree with standard usage in cosmology and  choose preferred observers that are at rest in space and their adapted tetrad axes point along the Cartesian coordinate directions
\begin{equation}\label{M2}
e^\mu{}_{\hat{\alpha}}=e^{-U}\,\delta^\mu_{\alpha}\,, \qquad     e_\mu{}^{\hat{\alpha}}=e^{U}\,\delta_\mu^{\alpha}\,.
\end{equation}
We seek an exact solution of the field Eqs.~\eqref{T9} and~\eqref{T10} in this case. 

The torsion tensor can be simply computed and is given by
\begin{equation}\label{M3}
 C_{\alpha \beta}{}^{\mu}=U_{\alpha}\, \delta^\mu_\beta- U_{\beta}\, \delta^\mu_\alpha\,, \quad  C_{\alpha \beta \gamma}=e^{2U}\,(U_{\alpha}\, \eta_{\beta \gamma}- U_{\beta}\, \eta_{\gamma \alpha})\,,
\end{equation}
where $U_{\mu}:= \partial_{\mu}U$ in our convention. Similarly,  the torsion vector, the contorsion tensor and the auxiliary torsion tensor are $C_{\alpha}=-3\,U_{\alpha}$,  $K_{\alpha \beta \gamma}=C_{\beta \gamma\alpha}$ and $ \mathfrak{C}_{\alpha \beta \gamma}=-2\,C_{\alpha \beta \gamma}$, respectively. Moreover, in this case, $\check{C}_\alpha=0$; therefore, 
\begin{equation}\label{M4}
 X_{\mu \nu \rho} = \mathfrak{C}_{\mu \nu \rho}\,. 
\end{equation}

The Einstein tensor for conformally flat spacetimes is given by~\cite{Ste}
\begin{equation}\label{M5}
^0G_{\mu \nu}=-2\,(U_{\mu \nu}-U_\mu\,U_\nu)+\eta_{\mu \nu}\eta^{\alpha \beta}(U_\alpha\,U_\beta+2\,U_{\alpha \beta})\,,
\end{equation}
where $U_{\mu \nu} = \partial_\mu \partial_\nu U$. We assume that the energy-momentum tensor of matter is given by a perfect fluid of energy density $\rho$ and pressure $P$ such that 
\begin{equation}\label{M6}
T_{\mu \nu}=\rho\,u_\mu\,u_\nu+P\,(g_{\mu \nu}+u_\mu\,u_\nu)\,,
\end{equation}
where $u^\mu$ is the 4-velocity vector of the perfect fluid. As in the standard cosmological models, we assume the fundamental particles are comoving with the preferred observers and are thus spatially at rest, namely, 
\begin{equation}\label{M7}
 u^\mu = e^{\mu}{}_{\hat 0} = e^{-U}\delta^\mu_0\,,  \qquad u_\mu = e^U\eta_{\mu 0}\,
\end{equation} 
and $\rho$ and $P$ are functions of conformal time $\eta$. 

Let us return to  the field Eqs.~\eqref{T9} and~\eqref{T10} and note that the constitutive relation~\eqref{N5} in this case reduces to
\begin{equation}\label{M8}
N_{\mu \rho \nu} = S \, \mathfrak{C}_{\mu \rho \nu} = -2S\, C_{\mu \rho \nu} = -2Se^{2U}(U_\mu \eta_{\rho \nu}-U_\rho\eta_{\mu \nu})\,.
\end{equation}
Therefore,
\begin{equation}\label{M9}
N_{\mu}{}^{\rho}{}_{\nu} = - 2S (U_\mu \delta^\rho_\nu-\eta^{\rho \alpha}U_\alpha \eta_{\mu \nu})\,, \qquad N_{\mu}{}^{\rho}{}_{\rho} = - 6S U_\mu\,.
\end{equation}
Using these relations in $Q_{\mu \nu}$ and $\mathcal{N}_{\mu \nu}$, 
\begin{equation}\label{M10}
Q_{\mu \nu}=U_\mu\,N_{\nu}{}^{\rho}{}_\rho-U_\rho\,N_{\nu}{}^{\rho}{}_\mu-\frac{1}{2}\,g_{\mu \nu}\,U_\alpha\,N^{\alpha \beta}{}_{\beta}\,, \quad \mathcal{N}_{\mu \nu}=e^{-U} \,\eta_{\nu \alpha}\,\frac{\partial}{\partial x^\beta}\,\Big(e^{3\,U}\,N^{\alpha \beta}{}_{\mu}\Big)\,,
\end{equation}
we find $Q_{\mu \nu}$  is symmetric 
\begin{equation}\label{M11}
Q_{\mu \nu}= - 4 S\,U_\mu U_\nu + S \,\eta_{\mu \nu}\,\eta^{\alpha \beta}U_\alpha U_\beta\,
\end{equation} 
 and
\begin{equation}\label{M12}
\mathcal{N}_{\mu \nu}= - 2 (S_\mu U_\nu -\eta_{\mu \nu}\,\eta^{\alpha \beta}S_\alpha U_\beta) - 2 S\,(U_\mu U_\nu -\eta_{\mu \nu}\,\eta^{\alpha \beta}U_\alpha U_\beta) - 2S(U_{\mu \nu} - \eta_{\mu \nu}\,\eta^{\alpha \beta}U_{\alpha \beta})\,.
\end{equation}
The second field Eq.~\eqref{T10} implies $ \mathcal{N}_{[\mu \nu]} = 0$, which means 
\begin{equation}\label{M13}
 S_\mu U_\nu = S_\nu U_\mu\,.
\end{equation} 
Hence, $dS \wedge dU = 0$. This equation has the natural solution that $dS$ is proportional to $dU$, namely,
\begin{equation}\label{M14}
 S = S(U)\,.
\end{equation}
The symmetric field Eq.~\eqref{T9} is simply Einstein's field equation together with a new source
\begin{equation}\label{M15}
Q_{\mu \nu} - \mathcal{N}_{\mu \nu} = - 2 \left(S -\frac{dS}{dU}\right)\,U_\mu U_\nu - \left(S +2\frac{dS}{dU}\right)\,\eta_{\mu \nu}\,\eta^{\alpha \beta}U_\alpha U_\beta + 2S(U_{\mu \nu} - \eta_{\mu \nu}\,\eta^{\alpha \beta}U_{\alpha \beta})\,.
\end{equation}
Let us write Eq.~\eqref{T9} in the form
 \begin{equation}\label{M16}
^{0}G_{\mu \nu} - Q_{\mu \nu} + \mathcal{N}_{\mu \nu} = \kappa T_{\mu \nu} - \Lambda g_{\mu \nu}\,.
\end{equation}
The left side of this equation can be written as
\begin{equation}\label{M17}
-2(1+S)(U_{\mu \nu} - \eta_{\mu \nu}\,\eta^{\alpha \beta}U_{\alpha \beta}) + 2 \left(1+S -\frac{dS}{dU}\right)\,U_\mu U_\nu + \left(1+S +2\frac{dS}{dU}\right)\,\eta_{\mu \nu}\,\eta^{\alpha \beta}U_\alpha U_\beta\,.
\end{equation}

To proceed, we note that in the context of standard \emph{flat} cosmology, the natural choice for $U(x)$ would be to assume $U = U(\eta)$. Therefore, $S = S(\eta)$ as well, in agreement with the time dependent nature of the background.  Moreover, we introduce the scale factor $a$, 
\begin{equation}\label{M18}
  e^U = a(\eta)\,, \qquad a':= \frac{da}{d\eta}\,, \qquad U_\mu = \frac{a'}{a}\delta^0_\mu\,.
\end{equation}
 With these assumptions, Eq.~\eqref{T9} has nonzero contributions for indices $(\mu ,\nu ) = (0 ,0)$ and $(\mu, \nu) = (i , j)$. We find
\begin{equation}\label{M19}
 \frac{3}{a^2}(1+S) \left( \frac{a'}{a}\right)^2 = \Lambda + 8\pi G \rho\, 
\end{equation} 
and 
\begin{equation}\label{M20}
 2(1+S) \left( \frac{a'}{a}\right)' + (1+S) \left( \frac{a'}{a}\right)^2 = a^2(\Lambda - 8\pi G P) - 2 \frac{dS}{dU} \left( \frac{a'}{a}\right)^2\,, 
\end{equation} 
respectively. When $S = 0$, we have the standard GR results, as expected. 

It is useful to express these equations in the more traditional form of standard flat cosmology. The solution of the modified TEGR field equations is the Cartesian tetrad field~\eqref{M2}, where $U$ is given by Eqs.~\eqref{M18}--\eqref{M20}. Henceforth, we employ the phrase ``Cartesian flat" for this model in order to indicate that the Euclidean three-dimensional spatial part of the spacetime metric of this model is indeed due to the Cartesian frame field that is the actual solution of the modified TEGR field equations.


\section{Modified Cartesian Flat Cosmology} 

In the traditional flat model, we assume a metric of the form
\begin{equation}\label{F1}
 ds^2 = -dt^2 + a^2(t)\,\delta_{ij}\,dx^i\,dx^j\,
\end{equation} 
in terms of \emph{cosmic time} $t$. With $dt = a d\eta$, the metric can then be written as $ds^2 = a^2 \eta_{\mu \nu}\, dx^\mu\,dx^\nu$, just as in the previous section. 

Let us write
\begin{equation}\label{F2}
 \dot{a} := \frac{da}{dt} = \frac{a'}{a}\,.
\end{equation} 
The traditional forms of the dynamical equations of our model become
\begin{equation}\label{F3}
 3(1+S) \left( \frac{\dot{a}}{a}\right)^2 = \Lambda + 8\pi G \rho\, 
\end{equation} 
and 
\begin{equation}\label{F4}
 2(1+S) \frac{\ddot{a}}{a} + (1+S) \left( \frac{\dot{a}}{a}\right)^2 = \Lambda - 8\pi G P - 2 \frac{dS}{dt} \frac{\dot{a}}{a}\,, 
\end{equation} 
respectively. Next, differentiating Eq.~\eqref{F3} with respect to cosmic time $t$ and using Eq.~\eqref{F4}, we find 
\begin{equation}\label{F5}
\frac{d\rho}{dt} = - 3(\rho + P) \frac{\dot{a}}{a} -  \frac{3}{8 \pi G} \frac{dS}{dt}\left( \frac{\dot{a}}{a}\right)^2\,,
\end{equation}
which implies, for $\rho + P\ge 0$ and $dS/dt >0$, that $\rho$ monotonically decreases as the universe expands. Indeed, we expect that as $t \to \infty$, the scale factor $a(t)$ approaches infinity  and $\rho$ and $P$ approach zero. 
Finally, it follows from Eqs.~\eqref{F3} and~\eqref{F5} that
\begin{equation}\label{F6}
\frac{d\rho}{dt} = - 3(\rho + P) \frac{\dot{a}}{a} - \left(\rho + \frac{\Lambda}{8 \pi G}\right)\, \frac{\dot{S}}{1+S}\,.
\end{equation}

It is important to recognize that in this modified Cartesian flat model the Cartesian tetrad frame field is such that the spatial part of the corresponding metric is the three-dimensional Euclidean space; moreover, it is possible to assume that $dS/dt \ne 0$. However, this is not the only solution of the field equations in this case.  In the next section, we show that the standard Friedmann-Lema\^itre-Robertson-Walker (FLRW) models are all solutions of the modified TEGR field equations; nevertheless, only  the modified Cartesian flat model described above has  proper physical significance.


\section{Modified FLRW Cosmology}

In our modified TEGR framework, it may be possible to extend the simple model of the previous section to the standard cosmological models. For the sake of completeness, we employ two simple but different representations of the FLRW models. 


\subsection{Isotropic Coordinates}

We begin with 
\begin{equation}\label{H1}
 ds^2 = -dt^2 + \frac{a^2(t)}{K^2(r)}\,\delta_{ij}\,dx^i\,dx^j\,, 
\end{equation} 
where $a(t)$ is now a characteristic radius, 
\begin{equation}\label{H2}
K(r)  = 1 + \frac{1}{4}\, k\, r^2\,, \qquad  r^2 = \delta_{ij}\, x^i x^j\,, 
\end{equation}  
and $k = 1, -1$, or $0$,  for the closed, open, or flat FLRW model, respectively. For metric~\eqref{H1}, the nonzero components of the  Einstein tensor are given by 
\begin{equation}\label{H2a}
^{0}G_{00} = 3\, \left(\frac{\dot{a}}{a}\right)^2 + 3\,\frac{k}{a^2}\,
\end{equation}
and
\begin{equation}\label{H2b}
^{0}G_{ij} = -\frac{1}{K^2}(2a\,\ddot{a} + \dot{a}^2 + k)\,\delta_{ij}\,.
\end{equation}

The fundamental observers are at rest in space and carry adapted tetrads that point along the coordinate directions. That is, 
\begin{equation}\label{H3}
e^\mu{}_{\hat 0}= \delta^\mu_{0}\,, \qquad   e^\mu{}_{\hat i}= \frac{K(r)}{a(t)}\delta^\mu_{i}\,
\end{equation}
and 
\begin{equation}\label{H4}
e_\mu{}^{\hat 0}= \delta_\mu^{0}\,, \qquad e_\mu{}^{\hat i}= \frac{a(t)}{K(r)}\delta_\mu^{i}\,.
\end{equation}
It is important to note that the tetrad frame field in this case is such that the corresponding spacetime metric has a spatial part that differs from flat Euclidean space for $k = \pm 1$. 

After some work, details of which we relegate to Appendix B, we find that the constraint equations imply $k\dot{S} = 0$. Therefore, the equations of modified TEGR are consistent with $S = S(t)$ in this case provided $k = 0$ and we thus recover the Cartesian flat model of the previous section. On the other hand, it is possible that $k \ne 0$, in which case $S$ must be constant. In the latter case, the main equations of FLRW model for for $k = \pm 1$ and  \emph{constant} $S$  become 
\begin{equation}\label{H5}
 \frac{3}{c^2}(1+S) \left( \frac{\dot{a}}{a}\right)^2 + 3(1+S)\,\frac{k}{a^2}  = \Lambda + \frac{8\pi G}{c^4} \rho\, 
\end{equation} 
and 
\begin{equation}\label{H6}
 \frac{2}{c^2}(1+S) \frac{\ddot{a}}{a} +\frac{1}{c^2} (1+S) \left( \frac{\dot{a}}{a}\right)^2 + (1+S)\,\frac{k}{a^2} = \Lambda - \frac{8\pi G}{c^4} P\,. 
\end{equation} 
For $S = 0$, we get the standard equations of FLRW model. 

In a spatially homogeneous and isotropic background that varies with time, we expect that the susceptibility $S$ would be time dependent as well, in close analogy with the electrodynamics of media.  That is, a constant $S$ would naturally correspond to a background that is independent of time, while a time-varying $S(t)$ would normally belong to a dynamic background.  It is therefore thought-provoking that for $\dot{S} \ne 0$ only the Cartesian flat model, i.e. with Cartesian tetrad frame and corresponding three-dimensional Euclidean space,  of standard cosmology is allowed within the framework of the modified theory of gravitation under consideration here.  In this connection, let us emphasize that the physical motivation for our particular modified TEGR is the close analogy with  the electrodynamics of media; indeed, the susceptibility $S(x)$ is associated with a spacetime in essentially the same sense that $\epsilon(x)$ and $\mu(x)$ are associated with a medium in electrodynamics. If the electromagnetic nature of the medium varies in space and time, we expect that such variation is reflected in $\epsilon(x)$ and $\mu(x)$. Of the standard dynamic FLRW models in our modified TEGR, the only one consistent with $dS/dt \ne 0$ is the Cartesian flat model. This circumstance thus provides a cogent justification for the Cartesian frame field and spatial flatness in our modified TEGR cosmology. 


\subsection{Spherical Coordinates}

In this case, the FLRW metric can be written in the form
\begin{equation}\label{h1}
 ds^2 = -dt^2 + a^2(t) [d\varpi^2 +f^2(\varpi) (d\vartheta^2 + \sin^2\vartheta\, d\varphi^2)]\,, 
\end{equation} 
where $f(\varpi) = \sin \varpi, \varpi$, or $\sinh \varpi$ according as $k = 1, 0$, or $-1$. Let us define $f'(\varpi) := df/d\varpi$ and so on; then, 
\begin{equation}\label{h2}
 f'' = -k f\,, \qquad f'^2 = 1- k f^2\,. 
\end{equation} 
 For metric~\eqref{h1}, the nonzero components of the  Einstein tensor are given by 
\begin{equation}\label{h3}
^{0}G_{00} = 3\, \left(\frac{\dot{a}}{a}\right)^2 + 3\,\frac{k}{a^2}\,
\end{equation}
and
\begin{equation}\label{h4}
^{0}G_{11} = -(2a\,\ddot{a} + \dot{a}^2 + k)\,\qquad ^{0}G_{22} =f^2\,^{0}G_{11}\,, \qquad ^{0}G_{33} = f^2 \sin^2\vartheta\,^{0}G_{11}\,.
\end{equation}

As before, the preferred observers in this case are spatially at rest. They carry adapted tetrads that point along the coordinate directions, namely, 
\begin{equation}\label{h5}
e^\mu{}_{\hat 0}= \delta^\mu_{0}\,, \quad   e^\mu{}_{\hat 1}= \frac{1}{a(t)}\delta^\mu_{1}\,,  \quad   e^\mu{}_{\hat 2}= \frac{1}{a(t)f(\varpi)}\delta^\mu_{2}\,, \quad e^\mu{}_{\hat 3}= \frac{1}{a(t)f(\varpi)\sin\vartheta}\delta^\mu_{3}\,
\end{equation}
and 
\begin{equation}\label{h6}
e_\mu{}^{\hat 0}= \delta_\mu^{0}\,, \quad e_\mu{}^{\hat 1}= a(t)\,\delta_\mu^{1}\,, \quad e_\mu{}^{\hat 2}= a(t)f(\varpi)\,\delta_\mu^{2}\,, \quad e_\mu{}^{\hat 3}= a(t)f(\varpi)\sin \vartheta \,\delta_\mu^{3}\,.
\end{equation}

Using these tetrads, the torsion tensor can be calculated. As in the case of isotropic coordinates, we find  $C_{\mu \nu 0} = 0$. The only nonzero components of the torsion tensor are given by
\begin{equation}\label{h7}
C_{011} = - C_{101} = a \dot{a}\,, \qquad C_{022} = - C_{202} = a \dot{a}f^2\,, \qquad C_{122} = - C_{212} = a^2 f f'\,
\end{equation} 
and
\begin{align}\label{h8}
\nonumber {}&C_{033} = - C_{303} = a \dot{a}f^2 \sin^2\vartheta\,, \qquad C_{233} = - C_{323} = a^2 f^2 \sin \vartheta \cos\vartheta\,, \\
{}& C_{313} = - C_{133} = -a^2 f f' \sin^2 \vartheta\,.
\end{align} 
Therefore, the torsion vector is given by
\begin{equation}\label{h9}
C_{0}  = -3\,\frac{\dot{a}}{a}\,, \qquad  C_{1} = -2 \frac{f'}{f}\,, \qquad C_{2} = -\frac{\cos \vartheta}{\sin \vartheta}\,, \qquad C_{3} = 0\,.
\end{equation}
 
As for the contorsion tensor,  we again find $K_{0\mu \nu} = 0$; likewise, its only nonzero components are given by
\begin{equation}\label{h10}
K_{101} = - K_{110} = a \dot{a}\,, \qquad K_{202} = - K_{220} = a \dot{a}f^2\,, \qquad K_{212} = - K_{221} = a^2 f f'\,
\end{equation} 
and
\begin{align}\label{h11}
K_{303} = - K_{330} = a \dot{a}f^2 \sin^2\vartheta\,, \qquad K_{331} = - K_{313} = -a^2 f f' \sin^2 \vartheta\,.
\end{align} 

The calculation of the  auxiliary torsion tensor using Eq.~\eqref{G2} is now straightforward. We find $\mathfrak{C}_{ij0} = 0$ and $\mathfrak{C}_{0i0} = - \mathfrak{C}_{i00} = C_i$. The other nonzero components of $\mathfrak{C}_{\mu \nu \rho}$  are given by
\begin{equation}\label{h12}
\mathfrak{C}_{011} = - \mathfrak{C}_{101} = -2a \dot{a}\,, \qquad \mathfrak{C}_{121} = - \mathfrak{C}_{211}  = a^2\, \frac{\cos \vartheta}{\sin \vartheta}\,,
\end{equation}
\begin{equation}\label{h13}
\mathfrak{C}_{022} = - \mathfrak{C}_{202} = -2a \dot{a} f^2\,, \qquad \mathfrak{C}_{122} = - \mathfrak{C}_{212}  = -a^2 f f'\,, 
\end{equation} 
\begin{equation}\label{h14}
\mathfrak{C}_{033} = - \mathfrak{C}_{303} = -2a \dot{a} f^2 \sin^2\vartheta\,, \qquad \mathfrak{C}_{313} = - \mathfrak{C}_{133}  = a^2 f f' \sin^2 \vartheta\,.
\end{equation}  

It is important to note that in the nonzero components of $C_{\mu \nu \rho}, K_{\mu \nu \rho}$, and $\mathfrak{C}_{\mu \nu \rho}$, the three indices cannot be all different. It follows that the torsion pseudovector vanishes in this case; that is, $\check{C} = 0$. The constitutive relation~\eqref{N5} thus reduces to $N_{\mu \nu \rho} = S\, \mathfrak{C}_{\mu \nu \rho}$. We are now in a position to calculate $Q_{\mu \nu}$, which is traceless and simply proportional to $S$, and $\mathcal{N}_{\mu \nu}$, which could involve derivative of $S$, namely,  $dS/dt$, since $S$ could only depend upon time in a dynamic spatially homogeneous spacetime. For the nonzero components of $Q_{\mu \nu}$, we find 
\begin{equation}\label{h15}
Q_{00} = -3\, S \left(\frac{\dot{a}}{a}\right)^2 - \frac{S}{a^2}\,\left(\frac{f'}{f}\right)^2\,,\qquad Q_{01} = -2 \,S \frac{\dot{a}}{a}\frac{f'}{f}\,,  \qquad Q_{02} = - S\frac{\dot{a}}{a}\frac{\cos \vartheta}{\sin\vartheta}\,,
\end{equation}
\begin{equation}\label{h16}
Q_{10} = 2\,Q_{01}\,, \qquad Q_{11} = -S \dot{a}^2 -S  \left(\frac{f'}{f}\right)^2\,, \qquad Q_{22} = -S \dot{a}^2 f^2\,, 
\end{equation}
\begin{equation}\label{h17}
Q_{20} = 2\,Q_{02}\,, \qquad  Q_{21}  =  - S\frac{f'}{f}\frac{\cos \vartheta}{\sin\vartheta}\,,   \qquad Q_{33} = -S \dot{a}^2 f^2\sin^2 \vartheta\,.    
\end{equation}

Similarly, for the nonzero components of $\mathcal{N}_{\mu \nu}$, we find
\begin{equation}\label{h18}
\mathcal{N}_{00} = \frac{S}{a^2}\left[3k -  \left(\frac{f'}{f}\right)^2\right]\,,  \qquad \mathcal{N}_{01} = Q_{01}  -2\,\frac{f'}{f} \frac{dS}{dt}\,, \qquad \mathcal{N}_{10} = Q_{10}\,,
\end{equation} 
\begin{equation}\label{h19}
\mathcal{N}_{11} = - \frac{S}{f^2}- 2\,\frac{d(a\dot{a}S)}{d t}\,, \qquad \mathcal{N}_{20} = 2\,Q_{02}\,, \qquad \mathcal{N}_{02} = Q_{02} - \frac{\cos \vartheta}{\sin\vartheta}\,\frac{dS}{dt}\,,
\end{equation}
\begin{equation}\label{h20}
\mathcal{N}_{21} = Q_{21}\,, \qquad \mathcal{N}_{22} = - kSf^2- 2\,f^2\,\frac{d(a\dot{a}S)}{d t}\,, \qquad \mathcal{N}_{33} = \mathcal{N}_{22}\sin^2 \vartheta\,.
\end{equation}

It is now straightforward to compute the components of $\mathbb{R}_{\mu \nu}  = Q_{\mu \nu} - \mathcal{N}_{\mu \nu}$ and show that the modified GR field equations~\eqref{T9} and~\eqref{T10} reduce in this case to
\begin{equation}\label{h21}
 \frac{3}{c^2}(1+S) \left( \frac{\dot{a}}{a}\right)^2 + 3(1+S)\,\frac{k}{a^2}  = \Lambda + \frac{8\pi G}{c^4} \rho\,, 
\end{equation}  
\begin{equation}\label{h22}
 \frac{2}{c^2}(1+S) \frac{\ddot{a}}{a} +\frac{1}{c^2} (1+S) \left( \frac{\dot{a}}{a}\right)^2 + (1+S)\,\frac{k}{a^2} +  \frac{2}{c^2}\,\frac{\dot{a}}{a}\,\frac{dS}{dt}  = \Lambda - \frac{8\pi G}{c^4} P\, 
\end{equation} 
and
\begin{equation}\label{h23}
\mathbb{R}_{01} = 2\,\frac{f'}{f} \frac{dS}{dt} = 0\,, \qquad \mathbb{R}_{02} = \frac{\cos \vartheta}{\sin\vartheta}\,\frac{dS}{dt} = 0\,.
\end{equation} 
The diagonal components of the modified GR field equations imply the familiar results~\eqref{h21}--\eqref{h22}, while the off-diagonal field equations given in Eq.~\eqref{h23} imply that $dS/dt = 0$ and hence $S$ is a constant for the closed, flat and open FLRW models. On the other hand, if we insist that $dS/dt \ne 0$ in Eq.~\eqref{h23}, then for tetrad frame field of the form~\eqref{h5}--\eqref{h6} with arbitrary $f(\varpi)$ we should have that $f$ is constant and $\cos \vartheta = 0$, or equivalently, $\sin \vartheta = 1$, in which case the frame field would become essentially Cartesian and the spatial part of metric~\eqref{h1} would become explicitly Euclidean. 

Modified TEGR, which is the local limit of NLG, is a tetrad theory. The tetrad formulation of GR that results in TEGR would work for any smooth tetrad that we adopt as our preferred frame field. However, NLG and modified TEGR remove this 6-fold degeneracy at each event in spacetime. The examples presented in this section in connection with modified TEGR demonstrate that the field equations in this case have a number of possible solutions. Among these, the only one that is physically acceptable is the modified Cartesian flat cosmological model which (a) requires the tetrad frame field to be Cartesian such that the spatial part of the spacetime metric is then Euclidean and (b) allows a susceptibility function $S(t)$ associated with the time-dependent dynamic spacetime such that $dS/dt \ne 0$. Similar results are obtained when we explore spatially homogeneous but anisotropic cosmology in the local limit of NLG~\cite{Tabatabaei:2023lec}.
  
It is interesting to explore some of the general consequences of our modified Cartesian flat model. This is the purpose of the next section. 


\section{Implications of the Modified Cartesian Flat Model}

To interpret the main equations of the modified Cartesian flat model physically, let us first recall the implications of the standard model with $S = 0$. In the standard benchmark model, the positive cosmological constant ($\Lambda > 0$) can be replaced with a perfect fluid source with $T_{\mu \nu} = P_\Lambda \,g_{\mu \nu}$, where  $P_{\Lambda} = - \rho_{\Lambda} = - \Lambda/(8 \pi G)$; that is, the cosmological constant represents dark energy with an equation of state parameter $w_{\Lambda} = -1$. As the universe expands, the energy densities of the matter and radiation monotonically decrease such that the cosmological constant eventually becomes the dominant source.  For $t \to \infty$, $\Lambda$ generates de Sitter's solution of GR. Indeed, in the standard model ($S = 0$), the universe asymptotically approaches a de Sitter phase of exponential expansion such that $a(t)$ increases as  $\exp{(\sqrt{\Lambda/3} \,t)}$. On the other hand, it is simple to demonstrate that de Sitter spacetime is not a solution of our modified Cartesian flat cosmological model so long as $dS/dt \ne 0$.  We prove this assertion by contradiction. 

Let us start with Eqs.~\eqref{F3} and~\eqref{F4} that govern the dynamics of the modified Cartesian flat model of cosmology and set $\rho = P = 0$. The result is  the system of ordinary differential equations
\begin{equation}\label{e1}
 3(1+S) \left( \frac{\dot{a}}{a}\right)^2 = \Lambda\, 
\end{equation}  
\begin{equation}\label{e2}
 2(1+S) \frac{\ddot{a}}{a} + (1+S) \left( \frac{\dot{a}}{a}\right)^2 = \Lambda - 2 \frac{dS}{dt} \frac{\dot{a}}{a}\,. 
\end{equation} 
For an expanding universe, Eq.~\eqref{e1} implies
\begin{equation}\label{e3}
\frac{\dot{a}}{a} = (\Lambda/3)^{1/2} (1+S)^{-1/2}\,. 
\end{equation}
Differentiating this equation with respect to time, we get
\begin{equation}\label{e4}
\frac{\ddot{a}}{a} -  \left( \frac{\dot{a}}{a}\right)^2 = - \frac{1}{2}(\Lambda/3)^{1/2} (1+S)^{-3/2}\,\frac{dS}{dt}\,. 
\end{equation}
Multiplying this relation by $2(1+S) > 0$ and subtracting the resulting equation from  Eq.~\eqref{e2}, we find
\begin{equation}\label{e4a}
 3(1+S) \left( \frac{\dot{a}}{a}\right)^2 = \Lambda - 2 \frac{dS}{dt} \frac{\dot{a}}{a} + (\Lambda/3)^{1/2} (1+S)^{-1/2}\,\frac{dS}{dt}\,,
\end{equation} 
which, in the light of Eq.~\eqref{e1},  implies
\begin{equation}\label{e4b}
2\frac{\dot{a}}{a}\frac{dS}{dt} = (\Lambda/3)^{1/2} (1+S)^{-1/2}\,\frac{dS}{dt}\,. 
\end{equation}
If  $dS/dt \ne 0$, we can divide both sides of Eq.~\eqref{e4b} by $dS/dt$; then, we are left with a relation that directly contradicts Eq.~\eqref{e3}. We conclude that de Sitter spacetime is not a solution of our modified Cartesian flat model.  This is consistent with the fact that de Sitter spacetime is not a solution of nonlocal gravity (NLG)~\cite{Mashhoon:2022ynk}. Consequently, the universe model under consideration here will never asymptotically approach a de Sitter phase. Henceforth, we assume $\Lambda = 0$. To incorporate dark energy in our model, we resort to a perfect fluid component with positive energy density $\rho_{de} > 0$ and negative pressure $P_{de} = w_{de} \,\rho_{de}$ such that 
$w_{de} < 0$. The implication of dynamic dark energy for cosmology has been further explored in~\cite{Tabatabaei:2023qxw}.

It is interesting to note that in Eq.~\eqref{F4}, the important additional term due to $dS/dt$ acts like added pressure. 
For $S= {\rm constant}$, the basic local thermodynamic relation for adiabatic processes, namely,  $d\mathbb{U} = - P\, d\mathbb{V}$ is satisfied. That is, imagine an amount of energy $\mathbb{U}$ of the background perfect fluid contained within a local sphere of radius $\ell$ that expands with the universe; then, with
\begin{equation}\label{f2}
 \mathbb{U} := \rho \mathbb{V}\,, \qquad \mathbb{V} = \frac{4\pi}{3} \ell^3\,,\qquad \frac{\dot \ell}{\ell} = \frac{\dot a }{a}\,,
\end{equation} 
we have 
\begin{equation}\label{f3}
\frac{d\rho}{dt} = - 3(\rho + P) \frac{\dot{a}}{a}\,. 
\end{equation} 
On the other hand, the variation in $S$ is related to variation in entropy $\mathbb{S}$.  More generally, the basic thermodynamic relation for a nonadiabatic process is
\begin{equation}\label{f4}
 d\mathbb{U} = \mathbb{T} d\mathbb{S} - P d\mathbb{V}\,, 
\end{equation} 
where $\mathbb{T}$ is the temperature and $\mathbb{S}$ is the entropy. Writing the change in heat as $\delta \mathbb{Q} := \mathbb{T} d\mathbb{S}$, we find in this case
\begin{equation}\label{f5}
\delta \mathbb{Q} = - \frac{1}{2G} (dS) a \dot{a}^2\,. 
\end{equation}
Usually,  heat is generated by friction. As the universe expands and $S$ increases, the corresponding entropy decreases. For a discussion of entropy variation in cosmology, 
see~\cite{Zeldovich:1983cr}. 

Within the context of standard cosmology, we expect that entropy increases as the universe expands. However, we deal here with comoving space and it can be shown that the local comoving entropy of matter and radiation remains constant in the standard model of cosmology and one can write the local law of energy conservation (or energy continuity equation) in the absence of heat flow. On the other hand, we find in our modified cosmology that there is more deceleration accompanied with negative entropy production. Normally, this should mean more order and less chaos. 

Let us write the main equations of the modified Cartesian flat model in terms of the Hubble ($H$) and deceleration ($q$) parameters that are defined as
\begin{equation}\label{f6}
 H := \frac{\dot{a}}{a}\,, \qquad qH^2 := - \frac{\ddot{a}}{a}\,.
\end{equation}  
Then, in the absence of the cosmological constant we have
\begin{equation}\label{f7}
 3(1+S) H^2 = 8\pi G \rho\, 
\end{equation}
and 
\begin{equation}\label{f8}
 3(1+S) q H^2 =  4\pi G (\rho + 3 P) + 3 \frac{dS}{dt}H\,. 
\end{equation}
It is clear from Eq.~\eqref{f8} that the extra term involving $dS/dt > 0$ contributes to the deceleration of the universe in the modified Cartesian flat model. 

To incorporate the accelerated expansion of the universe in the framework of the modified flat model, the dark energy component should be such that
\begin{equation}\label{e5}
4\pi G \rho_{de} (1+3w_{de}) + 3 \frac{dS}{dt}H < 0\,,
\end{equation}
in accordance with Eq.~\eqref{f8}. It follows from Eq.~\eqref{e5} that $1+ 3\, w_{de} < 0$ or $w_{de} < -1/3$. Dark energy in the modified theory is thus generally different from the way it is incorporated in the standard model. 

In terms of matter, radiation and dark energy components, the main equations of the modified Cartesian flat model, namely, Eqs.~\eqref{F3} and~\eqref{F4},  become
\begin{equation}\label{e6}
 3(1+S) \left(\frac{\dot{a}}{a}\right)^2 =  8\pi G \sum_i \rho_i\, 
\end{equation} 
and 
\begin{equation}\label{e7}
 2(1+S) \frac{\ddot{a}}{a} + (1+S) \left( \frac{\dot{a}}{a}\right)^2 =  - 8\pi G \sum_i P_i - 2 \frac{dS}{dt} \frac{\dot{a}}{a}\,, 
\end{equation} 
respectively. Here, as in the standard flat cosmological model, we have expressed the energy density and pressure  as the sum of the matter, electromagnetic radiation and dark energy content of the model, namely, 
\begin{equation}\label{e8}
\sum_i \rho_i = \rho_m + \rho_r + \rho_{de}\,, \qquad \sum_i P_i = P_m + P_r + P_{de}\,.  
\end{equation} 

As before, we can differentiate Eq.~\eqref{e6} with respect to cosmic time $t$ and combine the result with Eq.~\eqref{e7} to derive the analogue of Eq.~\eqref{F6} in this case, namely,
\begin{equation}\label{e9}
 \sum_i \frac{d\rho_i}{dt} + 3  \frac{\dot{a}}{a}\,\sum_i (\rho_i + P_i) + \frac{\dot{S}}{1+S}\,\sum_i\,\rho_i = 0\,.
\end{equation} 
Let us recall that in the standard model ($S = 0$), we have a consistent system of equations for the scale factor and it is sufficient to solve Eq.~\eqref{F3} with $S = 0$ to find $a(t)$; indeed, this comes about because the standard energy continuity equation is valid for each component.  The same kind of consistency can be achieved in the modified model if we assume that for each component the modified energy continuity equation holds, namely, 
\begin{equation}\label{e10}
  \frac{d\rho_i}{dt} + 3  \frac{\dot{a}}{a}\, (\rho_i + P_i) + \frac{\dot{S}}{1+S}\,\rho_i = 0\,. 
\end{equation} 
As usual, for each component the pressure is assumed to be proportional to the corresponding energy density with proportionality constant $w$.  Then, Eq.~\eqref{e10}  has the solution
\begin{equation}\label{e11}
\rho_i(t) = \rho_i(t_0)\frac{1+S(t_0)}{1+S(t)} \,a(t)^{-3(1+w_i)}\,, 
\end{equation} 
which for $S=0$ reduces to the result of the standard cosmological model. Here, $t = t_0$ represents the present epoch in cosmology and we assume $a(t_0) = 1$. We note that the restrictions on $w_{de}$ in the modified Cartesian flat model are different from those of the standard cosmological model; in particular, $w_{de} = -1$ is allowed, which implies
\begin{equation}\label{e5a}
\rho_{de}(t)|_{(w_{de} = -1)} = \rho_{de}(t_0) \frac{1+S(t_0)}{1+S(t)}\,.
\end{equation}

The main equation of the modified Cartesian flat model is thus obtained by substituting  Eq.~\eqref{e11}  in Eq.~\eqref{e6}, namely,
\begin{equation}\label{e12}
 3(1+S) \left(\frac{\dot{a}}{a}\right)^2 =  8\pi G\frac{1+S(t_0)}{1+S(t)} \sum_i \rho_i(t_0) \,a(t)^{-3(1+w_i)}\,.
\end{equation} 
To cast this equation in a form that would be similar to the standard flat cosmological model, let us define $\bar{\Omega}_i$ to be the ratio of $\rho_i(t_0)$ to a certain constant energy density $\rho_c$ given by
\begin{equation}\label{e13}
\rho_c := \frac{3 \bar{H}_0^2}{8\pi G}\,[1+S(t_0)]\,, 
\end{equation}
where $\bar{H}_0$ is the Hubble parameter of the modified model at the present epoch. Then, Eq.~\eqref{e12} at the present epoch can be written as  
\begin{equation}\label{e14}
\sum_i \bar{\Omega}_i = 1\,, \qquad \bar{\Omega}_i := \frac{\rho_i(t_0)}{\rho_c}\,
\end{equation}
and the main evolution equation of our modified Cartesian flat model can be expressed as
\begin{equation}\label{e15}
 \frac{1}{\bar{H}_0^2}\, \left(\frac{da}{dt}\right)^2 =  \left[\frac{1+S(t_0)}{1+S(t)}\right]^2 \left[\frac{\bar{\Omega}_m}{a}  + \frac{\bar{\Omega}_r}{a^2}  + \frac{\bar{\Omega}_{de}}{a^{(1+3w_{de})}}\right]\,.
\end{equation} 
This is the result that we need in order to confront the modified Cartesian flat cosmological model with observational data; that is, by a suitable comparison of the numerical solutions of this equation with current cosmological data, one should be able to choose an appropriate susceptibility function $S(t)$ and determine the energy content of the universe. In this process, the density parameters evolve differently due to $dS/dt > 0$ and the $\bar{\Omega}_i$ will in general turn out to be  different from the density parameters of the standard benchmark model. We emphasize that the density parameters $\bar{\Omega}_m$,  $\bar{\Omega}_r$ and $\bar{\Omega}_{de}$ in this equation refer to matter, radiation and dark energy, respectively; in particular, $\bar{\Omega}_m$ here refers to the total of the visible baryonic $\bar{\Omega}_b$ and effective cold dark matter $\bar{\Omega}_c$ contributions, i.e., $\bar{\Omega}_m=\bar{\Omega}_b+\bar{\Omega}_c$, at the present epoch.

We can interpret Eq.~\eqref{e15} as the total energy equation for a one-dimensional mechanical system. The net energy is zero, which is the sum of a positive effective kinetic energy part $(\mathbb{K}_{eff})$ plus a negative effective potential energy part $(\mathcal{V}_{eff})$, namely, 
\begin{equation}\label{e16}
\mathbb{K}_{eff}:=  \frac{1}{\bar{H}_0^2}\,\left[\frac{1+S(t)}{1+S(t_0)}\right]^2  \left(\frac{da}{dt}\right)^2\,
\end{equation}
and
\begin{equation}\label{e17}
\mathcal{V}_{eff}(a) := -\frac{\bar{\Omega}_m}{a}  - \frac{\bar{\Omega}_r}{a^2}  - \bar{\Omega}_{de}\,a^{-(1+3w_{de})}\,, \qquad  -(1+3w_{de}) >0\,,
\end{equation}
respectively. 
The general behavior of the negative effective potential energy is clear: its shape is roughly similar to an inverted harmonic oscillator potential; moreover,  near $a = 0$, it is dominated by $-1/a^2$ due to radiation, while near $a = \infty$, the dark energy component dominates. In this way, we can illustrate the general behavior of $a(t)$: Near the $t = 0$ singularity where $a = 0$, the universe expands very rapidly until $\mathcal{V}_{eff}$ approaches its maximum value where the expansion slows and there is a loitering phase around the maximum of the effective potential after which the universe expands rapidly again and may go through an accelerating phase depending on the strength of the dark energy component that could render $q$ negative. The important point here is that this general picture is independent of the form of $S(t)$. However, the details do depend on $S(t)$, especially where the expansion of the universe might actually accelerate whenever $q < 0$, namely,  
\begin{equation}\label{e18}
 (1+3w_{de})\bar{H}_0^2\,\bar{\Omega}_{de}\,\frac{[1+S(t_0)]^2}{1+S(t)} \,a(t)^{-3(1+w_{de})} + 2\, \frac{dS}{dt}H < 0\,.
\end{equation}

To illustrate the nature of our main result, let us assume our cosmology contains only one component with an equation of state parameter $w > -1$; that is, $\rho_w(t_0) = \rho_c$ and $\bar{\Omega}_w = 1$.  Then, Eq.~\eqref{e15} reduces to a simple ordinary differential equation that can be easily solved and the solution is
\begin{equation}\label{e19}
 a(t) = \left[\frac{\xi(t)}{\xi(t_0)}\right]^{2/[3(1+w)]}\,, \qquad  \xi(t) :=  \int_0^t\, \frac{dt}{1+S}\,.
\end{equation}
The Hubble ($H$) and deceleration ($q$) parameters can be simply computed in this case and we find 
\begin{equation}\label{e20}
H(t) = \frac{2}{3(1+w) (1+S) \xi(t)}\,, \qquad  q(t) = \frac{1}{2}(1+3w) + \frac{3}{2}(1+w) \frac{dS}{dt} \xi(t)\,.
\end{equation}
In the case of the standard model $(S = 0)$ for matter $(w = 0)$, $\xi(t) = t$ and we recover the  Einstein--de Sitter model with $a = (t/t_0)^{2/3}$, $H = 2/(3t)$ and $q = 1/2$; for dark energy, on the other hand, $-1< w_{de} < -1/3$ and $q(t)$ should be negative during accelerated epochs of the expansion of the universe. For dark energy with $w_{de} = -1$, we find 
\begin{equation}\label{e21}
H(t)|_{(w_{de} = -1)} = H(t_0) \frac{1+S(t_0)}{1+S(t)}\,,
\end{equation}
just as in Eq.~\eqref{e5a}. Finally, the case of dark energy with $w_{de} < -1$ is discussed in Appendix C. 

To go forward, we must compare the predictions of the modified Cartesian flat model with observational data. We make a beginning in this direction in the next section; however, a more complete analysis necessitates the investigation of linear perturbations of our modified Cartesian flat model~\cite{TaBa}. 


\section{Modified Cartesian Flat Model: $H_{\rm MTEGR}(z)$ and $H_0$ Tension}

Within the $\Lambda$CDM framework, the standard benchmark cosmological model with $S = 0$ has had significant success in explaining the vast majority of observations in connection with the cosmic microwave background (CMB)~\cite{Planck:2018vyg} and large scale structure formation~\cite{BOSS:2016wmc}. However, in recent years some discrepancies have been observed~\cite{Perivolaropoulos:2021jda, DAgostino:2023cgx}. These discrepancies could be hints for physics beyond the standard benchmark model of cosmology. The most recent discrepancy that is now under frequent discussion is related to the measurement of $H_0$, the relative expansion rate of the universe at the present time~\cite{Riess:2020fzl}. 
There is a 4--5 $\sigma$ discrepancy between the measurement of the Hubble constant using local studies of the nearby supernovas, for instance, and the measurement of the recession rate using the CMB on the basis of the $\Lambda$CDM model~\cite{DiValentino:2021izs}. This inconsistency has opened up a new arena in cosmological studies. Many alternative models have been suggested to reconcile this tension, from the early dark energy models to late-time modified gravity theories and so on~\cite{DiValentino:2021izs, Khosravi:2017hfi, Nygaard:2023gel, McConville:2023xav}.

To examine in detail the observational consequences of the modified Cartesian flat model, we must integrate the ordinary differential Eq.~\eqref{e15} to determine the temporal evolution of this universe model and then compare the results with cosmological data. For this purpose, we need a separate extensive investigation involving models of structure formation that is beyond the scope of the present work; in this connection, the linear perturbation regime has been studied in detail and is contained in~\cite{TaBa}. On the other hand, to indicate that the Cartesian flat model has the potential to help in the resolution of the $H_0$ tension, we employ in the present work observational data that are essentially geometric in nature known as the distance-redshift probes and do not involve cosmological perturbations. 

To this end, we can regard Eq.~\eqref{e15} as an \emph{algebraic} relation that expresses the Hubble parameter $H$ as a function of the cosmological redshift $z$, where
\begin{equation}\label{Q1}
 z := \frac{1}{a(t)} -1\,.
\end{equation}
Let us note that for the standard benchmark model
\begin{equation}\label{Q2}
H_{\Lambda \rm CDM} = H_0\, \Big[\sum_i \Omega_i \,(1+z)^{3(1+w_i)}\Big]^{1/2}\,,
\end{equation} 
while the Hubble parameter $H_{\rm MTEGR}(z)$ for our modified Cartesian flat TEGR model is given by Eq.~\eqref{e15}, namely, 
\begin{equation}\label{Q3}
H_{\rm MTEGR} = \bar{H}_0\, \Gamma(z)\Big[\sum_i \bar{\Omega}_i \,(1+z)^{3(1+w_i)}\Big]^{1/2}\,.
\end{equation} 
Here, we have introduced
\begin{equation}\label{Q4}
\Gamma(z) := \frac{1+S(0)}{1+S(z)}\,.
\end{equation} 
The function $\Gamma(z)$ is such that $\Gamma(0) = 1$, so that for the nearby universe, $0 \le z \ll 1$, the modified Cartesian flat model is not substantially different from the standard benchmark model. This implies that to solve the problem of $H_0$ tension, we must resort to cosmological data in the early universe. 

To proceed, we recall that the susceptibility function $S(t)$, which can be expressed as a function of the scale factor $a(t)$,  must be determined on the basis of observation. 
We therefore assume that $S(t)$ is proportional to $a(t)$ to some power; that is, we adopt the simple expression 
\begin{equation}\label{Q5}
 S(z) = \alpha (1+z)^\beta\,, \qquad \alpha > 0 \,, \qquad \beta < 0\,,
\end{equation} 
where $\alpha$ and $\beta$ are constant parameters that should be determined from observational data. Thus $S(z)$ is a monotonically decreasing function of $z$; in fact, $S$ is zero at $z = \infty$ and increases monotonically to a positive constant $\alpha$ at the present epoch ($z = 0$); moreover, $\Gamma(z)$ is an increasing function of $z$, since $S(z)$ monotonically decreases with $z$.

We need to compare observational data near the decoupling epoch with the predictions of Eqs.~\eqref{Q3}--\eqref{Q5}. To this end, we choose a cosmological standard candle, namely, the dimensionless number $\theta(z)$, which is the angle of view of the sound horizon.  For a data point at redshift $z$, this can be calculated as
\begin{equation}\label{Q6}
\theta(z) = \frac{r_d}{\mathcal{D}(z)}\,,
\end{equation}
where $r_d$ is the sound horizon at the drag epoch at which the baryons were free from the Compton drag of photons~\cite{Hu:1995en,Eisenstein:1997ik}
\begin{equation}\label{Q7}
r_d := \int_{z_d}^{\infty} \frac{c_s(z')}{H(z')} dz'\,
\end{equation} 
and $\mathcal{D}(z)$ is the distance to the observed structure
\begin{equation}\label{Q8}
\mathcal{D}(z) := \int_{0}^{z} \frac{c}{H(z')} dz'\,.
\end{equation}
The Hubble parameter $H$ in the calculation of $r_d$ is mostly dominated by the contribution of radiation, while the $H$ in $\mathcal{D}$ is mainly calculated in the matter dominated era. 
Here, the sound speed is given by
\begin{equation}\label{Q9}
c_s(z) = c\,\left[3\left(1+\frac{3\hat{\Omega}_b(z)}{4\hat{\Omega}_r(z)}\right)\right]^{-1/2}\,,\qquad \hat{\Omega}_i = \Omega_i \,(1+z)^{3(1+w_i)}\,.
\end{equation}

Before calculating these quantities for our modified Cartesian flat model, let us note that for the standard flat $\Lambda$CDM model according to~\cite{Planck:2018vyg}, $r_d \simeq 147.21 \pm 0.48$ Mpc is the comoving sound horizon at the end of the baryonic-drag epoch corresponding to redshift $z_d\simeq 1059.39 \pm 0.46$. Furthermore, the photon density parameter $\hat{\Omega}_r$ in Eq.~\eqref{Q9} is fixed by CMB temperature observations~\cite{Fixsen:2009ug} and baryon density parameter $\hat{\Omega}_b$ is fixed by the big-bang nucleosynthesis~\cite{Pisanti:2007hk}. The dark energy  is determined by the cosmological constant in the standard model; therefore, $\Omega_{\Lambda}=1-\Omega_c-\Omega_b-\Omega_r$ for the standard flat  $\Lambda$CDM model and $w_{\Lambda}=-1$.

The corrections that the local limit of NLG imposes on the above equation for $\theta$ can now be simply derived by using Eqs.~\eqref{Q3}--\eqref{Q5}. That is, 
\begin{equation}\label{Q10}
r_d|_{\rm MTEGR} = \int_{z_d}^{\infty} \frac{c_s(z')}{H_{\rm MTEGR}(z')} dz'\,
\end{equation}  
\begin{equation}\label{Q11}
\mathcal{D}(z)|_{\rm MTEGR} = \int_{0}^{z} \frac{c}{H_{\rm MTEGR}(z')} dz'\,
\end{equation}
and
\begin{equation}\label{Q12}
\theta|_{\rm MTEGR} =  \frac{r_d|_{\rm MTEGR}}{\mathcal{D}(z)|_{\rm MTEGR}}\,.
\end{equation}
As described below, we calculate the theoretical value of $\theta$ from this formula and compare the results with our observed values of $\theta$ and conclude that the early value of $H_0$ could possibly be larger and could resolve  the $H_0$ tension. In particular,  we note that $\bar{\Omega}_r = \Omega_r$ and $\bar{\Omega}_b = \Omega_b$, as these are fixed by the CMB observations and big-bang nucleosynthesis, respectively~\cite{Pisanti:2007hk,Fixsen:2009ug}. Consequently, $c_s(z)$ remains the same as in the standard model. We consider $\bar{\Omega}_c$ as a parameter and calculate  $\bar{\Omega}_{de} =1-\bar{\Omega}_c- \bar{\Omega}_b- \bar{\Omega}_r$ from the flatness condition. Moreover, we select $w_{de} = - 1$; this choice, as discussed in connection with Eq.~\eqref{e5a}, is permitted in the modified Cartesian flat model, and it also happens to be preferred when we treat $w_{de}$ as a free parameter in our numerical experiments.   In this way, we are left with a  4-dimensional parameter space consisting of $\alpha$, $\beta$, $\bar{H}_0$ and $\bar{\Omega}_{c}$.  Let us note that $\bar{H}_0$ appears as a coefficient in Eq.~\eqref{Q3} and subsequently  drops out in calculating the ratio $\theta|_{\rm MTEGR}$ in Eq.~\eqref{Q12}; nevertheless, the Hubble parameter is involved in calculating $\theta|_{\rm MTEGR}$ since $\bar{\Omega}_r\, \bar{H}_0^2$ is constrained by the CMB temperature~\cite{Fixsen:2009ug}. 

Within our 4-dimensional parameter space, we compute $\theta|_{\rm MTEGR}$ and compare our result with observational data. To analyze the data, we employ the Monte Carlo Markov Chain (MCMC) algorithm to look within the space of parameters for the best-fitting values by means of the $\chi^2$ statistic.


\subsection{Numerical Analysis}
	
The main purpose of our numerical work is to use some selected cosmological observations to constrain $\bar{H}_0$, $\bar{\Omega}_c$ and $S(z)$. The dataset consists of 16 observations; of these,  15 are  data points for the Baryon Acoustic Oscillations (BAO) in large-scale structures~\cite{Bassett:2009mm}, and the remaining  datum is related to the location of the first peak in the angular power spectrum of CMB temperature anisotropies. The CMB first peak position is a distance-redshift probe and will henceforth be referred to as ``CMB" for the sake of brevity. We employ CMB data from Planck~\cite{Planck:2018vyg},  BAO data from 6dF galaxy survey~\cite{Beutler:2011hx}, dark energy survey (DES)~\cite{DES:2017rfo}, main galaxy sample (MGS) from SDSS DR7~\cite{Ross:2014qpa}, WiggleZ~\cite{Kazin:2014qga}, luminous red galaxy (LRG) from SDSS DR14~\cite{Bautista:2017wwp}, Ly$\alpha$~\cite{duMasdesBourboux:2017mrl, Bautista:2017zgn}, BOSSDR12~\cite{BOSS:2016wmc} and SDSS DR14 quasar surveys~\cite{Zarrouk:2018vwy}. We name this dataset CMB+BAO in our plots and table.  

\begin{figure}[h]
\centering
\includegraphics[scale=0.7]{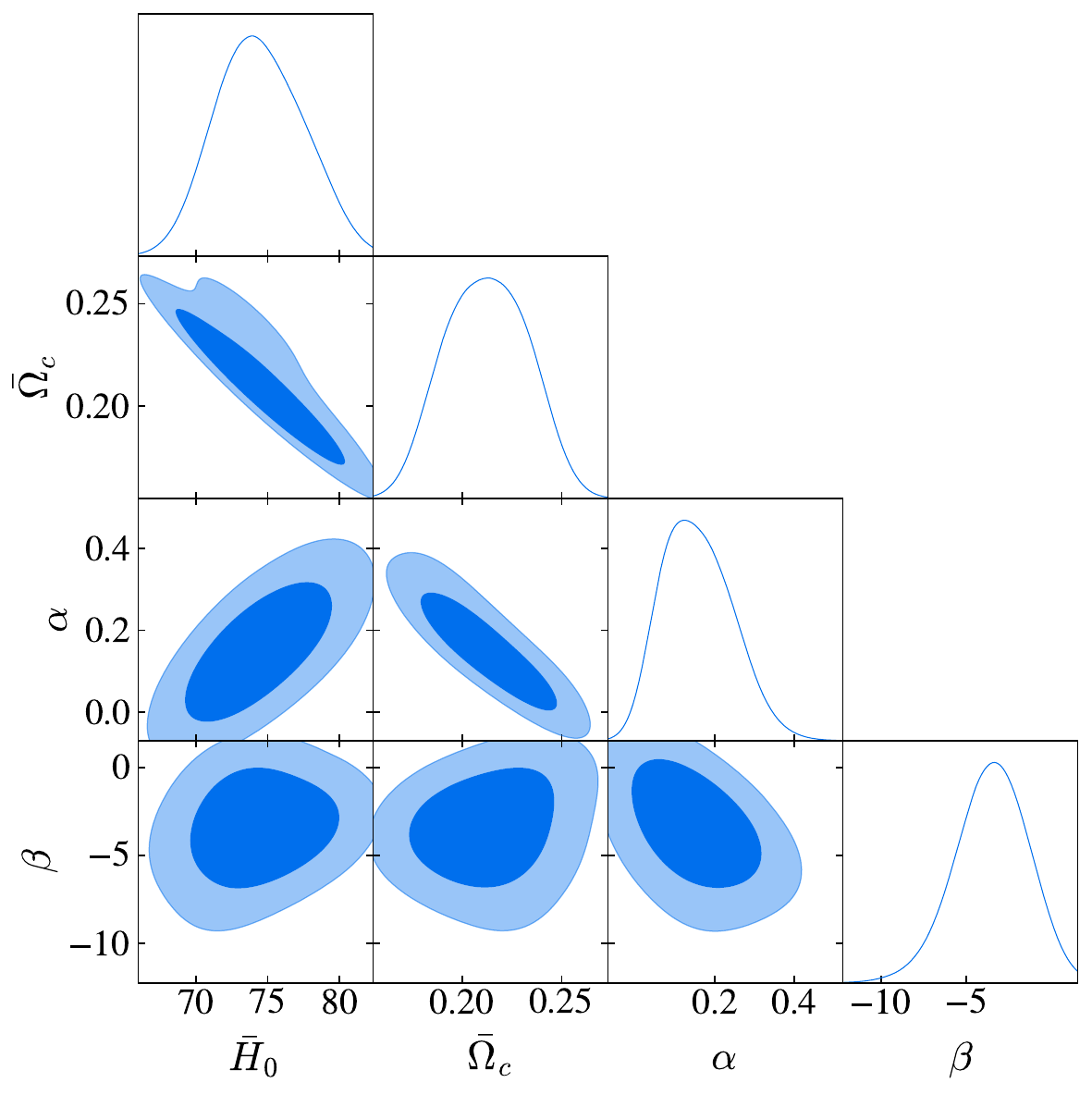}
\caption{Triangle plot showing the two-dimensional posterior probability results by means of the MCMC algorithm using  CMB+BAO dataset  for the MTEGR model with $S(z)=\alpha(1+z)^{\beta}$ ansatz. This plot is produced by employing the publicly available code \texttt{GetDist} \cite{Lewis:2019xzd}. The  smaller dark ( larger light) shaded regions show the 1$\sigma$  (2$\sigma$) confidence levels.}\label{fig:triangelPlot}
\end{figure}

The simple choice of $S(z)=\alpha (1+z)^\beta$ adds two new parameters $\alpha$ and $\beta$ to the set of parameters that we need to constrain using our data points.
Accordingly, the parameter set is: $\alpha$, $\beta$, $\bar{H}_0$ and $\bar{\Omega}_{c}$ for our modified Cartesian flat TEGR model (MTEGR). We search this 4-dimensional parameter space via the MCMC algorithm for the best-fitting parameters that minimize $\chi^2$ and match the theory with data. The result is given in the posterior probability diagram in Figure~\ref{fig:triangelPlot}. While we do not impose  prior conditions on the values of $\alpha$ and $\beta$, our results  are in alignment with  $\alpha > 0$ and $\beta<0$, which demonstrates the internal consistency of our model.  
 
\begin{figure}[h]
\centering
\includegraphics{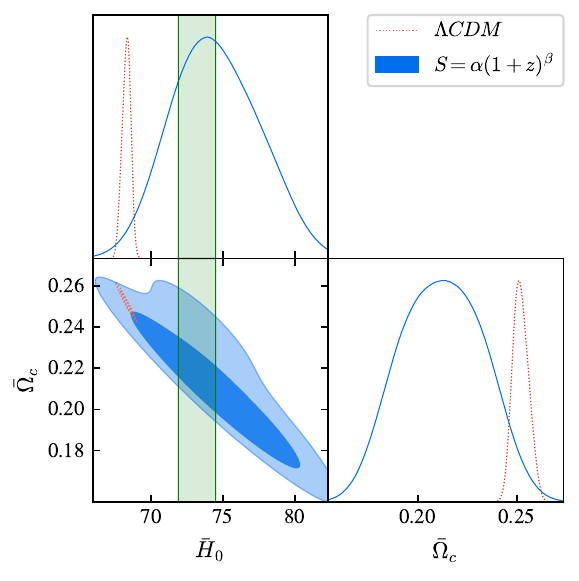}
\caption{Comparison of the posterior probability results of our MTEGR model (shaded oval contours and solid curves) with the $\Lambda$CDM model (dotted contour and curves) using the CMB+BAO dataset. This plot is produced by employing the publicly available code \texttt{GetDist} \cite{Lewis:2019xzd}. The shaded rectangular band represents the direct measurement of $H_0$ from the low-redshift observations by Riess \textit{et al.}~\cite{Riess:2019cxk}. The oval shaded regions show the 1$\sigma$ and 2$\sigma$ confidence levels.}\label{fig:comparePlot}
\end{figure}

In Figure~\ref{fig:comparePlot}, we plot the confidence levels regarding matter density and Hubble parameter. The shaded rectangular band represents the direct measurement of $H_0$ from the low-redshift observations by Riess \textit{et al.}~\cite{Riess:2019cxk}. Hereafter, we refer to the work of Riess \textit{et al.}~\cite{Riess:2019cxk} as ``R19". The dotted contour and  curves show the confidence levels for $\Omega_c$ and $H_0$ in the $\Lambda$CDM model; indeed, in the latter case, the dotted contour and curve do not touch the rectangular band, demonstrating $H_0$ tension. On the other hand, the solid curve and oval shaded regions touch the rectangular band, which means that our modified TEGR model can resolve the $H_0$ tension. This circumstance comes about due to the new parameters $\alpha$ and $\beta$ that characterize the susceptibility function $S$ in our model. Indeed, for $\alpha = 0$, $S = 0$ and we recover the standard GR cosmological model. We find that $\alpha$ is positively correlated with $H_0$; that is, $\alpha \approx 0.2$ implies a larger value for $H_0$, which relaxes the tension.  

Once it becomes clear that our model can alleviate the $H_0$ tension, we can  include Hubble constant data from the local universe by Riess \textit{et al.}~\cite{Riess:2019cxk} and repeat the $\chi^2$ analysis. In Table~\ref{table:1}, we show the best results for $\Lambda$CDM and MTEGR models with the two data sets, namely, CMB+BAO data and CMB+BAO+R19 data. To find a model that fits the observational data, we employ the MCMC algorithm to minimize the value of $\chi^2$ in order to find the best-fitting parameters for the model. To estimate the quality of fit, we employ the Akaike information criterion (AIC)~\cite{Akaike:1974} and Bayesian information criterion (BIC)~\cite{Schwarz:1978tpv}, where $\text{AIC} = 2k - 2 \ln(\mathcal{L_{\text{max}}})$ and $\text{BIC} = k\,\ln(n) - 2 \ln(\mathcal{L_{\text{max}}})$, respectively. Here, $n$ is the number of observational data points, $k$ is the number of fitting parameters and $\mathcal{L_{\text{max}}}$ is the maximized likelihood achieved by our MCMC algorithm. The likelihood is proportional to $\exp(-\chi^2/2)$ for Gaussian variables. Model selection criteria AIC and BIC impose different penalties when the number of parameters $k$ is increased. In Table~\ref{table:1}, $n = 16$ for the CMB+BAO data, while $n = 17$ for the CMB+BAO+R19 data; moreover, $k = 4$ for our model (MTEGR), while $k = 2$ for the $\Lambda$CDM model. The values of AIC and BIC in the right subtable demonstrate that our model is preferred for describing the CMB+BAO+R19 dataset. The simple algebraic approach presented here turns out to be consistent with the detailed analysis of cosmological perturbations where the full CMB data are employed~\cite{TaBa}.

\begin{table}[h!]
\centering
\subfloat[]{
\begin{tabular}{c | c  c } 
\hline\hline
Data & ~~~~CMB+BAO  \\ 
\hline\hline
Models & 	MTEGR &$\Lambda$CDM   \\ 
\hline
$h$& $0.74\pm 0.03$ & $0.683 \pm 0.005$ \\ 
$\Omega_c$& $0.21\pm0.02$ & $0.252\pm0.004$ \\ 
$\alpha$& $0.2\pm0.1$ & $\_$ \\
$\beta$ &$-4\pm2$ & $\_ $  \\
\hline  
$\chi^2$& $18.8$ & $21.0$  \\
AIC & $19.6$ & $17.8$  \\
BIC & $22.7$ & $19.4$  \\ [1ex] 
\hline\hline
\end{tabular}
}
\quad \quad
\subfloat[]{
\begin{tabular}{c | c  c } 
\hline\hline
Data &~~~~CMB+BAO+R19 \\ 
\hline\hline
Models & 	MTEGR &$\Lambda$CDM   \\ 
\hline
$h$& $0.73\pm 0.01$ & $0.686 \pm 0.003$ \\ 
$\Omega_c$& $0.22\pm0.01$ & $0.248\pm0.004$ \\ 
$\alpha$& $0.16\pm0.06$ & $\_$ \\
$\beta$ &$-5\pm2$ & $\_ $  \\
\hline  
	
$\chi^2$& $19.1$ & $30.0$  \\
AIC & $22.7$ & $29.6$  \\
BIC & $26.0$ & $31.3$  \\  [1ex] 
\hline\hline
\end{tabular}
}
\caption{Results from the MCMC algorithm for both MTEGR and $\Lambda$CDM models. In the left subtable (a), we use the CMB+BAO dataset.  In the right subtable (b), we use the CMB+BAO+R19 dataset. Note that $h := H_0 / (100~ \text{km/s/Mpc})$. }
\label{table:1}
\end{table}


\section{Discussion}

We have investigated the local limit of nonlocal gravity (NLG), which can help elucidate the nature of NLG in close analogy with the electrodynamics of media. Within the framework of locally modified TEGR (i.e., the local limit of NLG), the extension of the standard FLRW cosmological models has been worked out. The most sensible and compelling case involves the modified Cartesian flat  model, which has therefore been studied in some detail in the present work.  Other than matter and radiation, the modified Cartesian flat model involves a dark energy component that is different from a positive cosmological constant, since de Sitter spacetime is not a solution of the locally modified Cartesian flat model. 

The constitutive relation of NLG permits a different cosmic expansion history as compared to the prediction of the standard $\Lambda$CDM benchmark model~\cite{Chicone:2017oqt, Chicone:2015sda}. A similar outcome is expected for the  local limit of nonlocal gravity. Consequently, the locally modified TEGR cosmological models can give rise to a rich phenomenology to be confronted with observational data. In particular, the locally modified TEGR predicts a dynamic dark energy that under linear perturbations implies a different equation for the evolution of dark matter density contrast~\cite{TaBa}. Therefore, standard structure formation and related observational parameters will be altered in this model. Current observational programs to determine the equation of state of dark energy involve SNeIa, weak lensing, BAO and cluster counts~\cite{Zhao:2017cud}. In this connection, it is worth mentioning that one of the main aims of the ongoing and future large-scale structure probes like DESI~\cite{DESI:2018ymu}, Euclid, Nancy Grace Roman Space Telescope, and Vera Rubin Observatory~\cite{Rose:2021irr} is to constrain the equation of state of dark energy to determine whether it is consistent with a cosmological constant. 

In connection with recent tensions in cosmology, we note that there is current interest in quantum-inspired nonlocal cosmological models~\cite{Bouche:2023xjw, Giani:2023tai} as well as in various  modified teleparallel gravity theories~\cite{Saridakis:2023pzo, Mandal:2023bzo}. 

It should be mentioned that besides the $H_0$ tension, there is a less significant discrepancy known as $\sigma_8$ tension.  The problem is related to cosmological observations that suggest a smaller amplitude in the matter perturbation in the late-time universe as compared to the prediction of $\Lambda$CDM  on the basis of CMB data. The late-time observations that show this tension are mainly in the domain of weak lensing results~\cite{Joudaki:2016kym, Hildebrandt:2018yau, KiDS:2020suj} and growth-rate measurements~\cite{Perivolaropoulos:2021jda}. The fading memory effect of NLG for all components of the universe can be a hint that the modified TEGR model can address these tensions as well. 


\section*{ACKNOWLEDGMENTS}

S. B. has been partially supported by the Abdus Salam International Centre for Theoretical Physics (ICTP) under the junior associateship scheme.  J. T. thanks Abdolali Banihashemi for helpful discussions regarding the analysis of  cosmological data. B. M. is grateful to Yuri Obukhov for valuable discussions. 

\appendix

\section{Eikonal Approximation for Wave Eq.~\eqref{W6}}

The general connection between the wave equation and the motion of classical particles in spacetime is established via the eikonal (``WKB") approximation method. In the present case, we are interested in the high-frequency solution of 
\begin{equation}\label{A1}
 (1+S)\,\square\,\bar{h}^{kl} - \frac{dS}{dt} \frac{\partial \bar{h}^{kl}}{\partial t} = 0\,,
\end{equation}
where $S$ is a real function of time $t$.  For the solution of the wave equation, we assume an eikonal  expansion of the form
\begin{equation}\label{A2}
\bar{h}^{kl} = e^{i \, \mathcal{S}(x)/\lambdabar} \sum_{j = 0}^{\infty} \lambdabar^j \mathbb{F}^{kl}_j(x)\,,
 \end{equation}
where the action $ \mathcal{S}(x)$ is a \emph{real} scalar function of the spacetime coordinates and the reduced wavelength $\lambdabar = c/\omega$ approaches zero. In the asymptotic series in Eq.~\eqref{A2}, $\mathbb{F}^{kl}_j$, $j = 0, 1, 2, ...$,  are slowly varying functions such that  $\mathbb{F}^{kl}_0 \ne 0$ by assumption. The wave function $\bar{h}^{kl}$ represents a real gravitational wave and the wave equation~\eqref{A1} is linear; therefore, we can simply deal with $\bar{h}^{kl}$ as a complex quantity with the proviso that its real part has physical significance. 

The substitution of ansatz~\eqref{A2} in the second-order wave equation~\eqref{A1} results in an expansion in increasing powers of $\lambdabar$ beginning with $\lambdabar^{-2}$. The equation of motion in the WKB approximation is obtained by setting the coefficients of $\lambdabar^{-2}$ and $\lambdabar^{-1}$ terms equal to zero. We note that the wavelength of the radiation is not a scalar quantity and the traditional eikonal approach is not generally covariant. It turns out, however, that the covariant eikonal approximation is physically meaningful only in the actual limit of zero wavelength~\cite{Mash87}. 

Substituting ansatz~\eqref{A2}  in the wave equation, we find that the vanishing of the coefficient of $\lambdabar^{-2}$ term results in
\begin{equation}\label{A3}
 \eta^{\mu \nu}  \mathcal{S}_{,\mu} \, \mathcal{S}_{,\nu} = 0\,,
\end{equation} 
since $\mathbb{F}^{kl}_0$ does not vanish by assumption. Thus the radiation follows a straight null path in the eikonal limit and for the solution of Eq.~\eqref{A3} we can write
\begin{equation}\label{A4}
 \mathcal{S} =  \mathcal{S}_0 ( t - \mathbf{n} \cdot \mathbf{x} )\,,
\end{equation} 
where  $\mathbf{n}$ is a constant unit vector and $\mathcal{S}_0$ is a nonzero constant. 

Next, the vanishing of the coefficient of $\lambdabar^{-1}$ term results in 
\begin{equation}\label{A5}
2 (1+S) \frac{\partial \mathbb{F}^{kl}_0}{\partial t}  + 2 (1+S)n^i \frac{\partial \mathbb{F}^{kl}_0}{\partial x^i} +  \frac{dS}{dt} \, \mathbb{F}^{kl}_{0} = 0\,,
\end{equation}
once we take Eq.~\eqref{A4} into account. Moreover, the wave function satisfies the transverse gauge condition $\bar{h}^{kl}{}_{,l}=0$, which in this case simply means $\mathbb{F}^{kl}_{0}\, n_l = 0$.

Let $\mathbf{e}_1$ and $\mathbf{e}_2$, $\mathbf{e}_1 \cdot \mathbf{e}_2 = 0$, be constant unit vectors that are orthogonal to the direction of propagation of the null ray $\mathbf{n}$; moreover, let 
$\mathbf{e}$ be a linear combination of $\mathbf{e}_1$ and $\mathbf{e}_2$.   Then, the solution of  Eq.~\eqref{A5} can be written as
\begin{equation}\label{A6}
\mathbb{F}^{kl}_{0}(t, \mathbf{x}) = [f_{11}\, e_1^k e_1^l + f_{12}\, (e_1^k e_2^l +  e_2^k e_1^l) + f_{22}\, e_2^k e_2^l]\, \mathcal{F}(\mathbf{e} \cdot \mathbf{x})\,[1+S(t)]^{-1/2}\,,
\end{equation}
where $f_{11}$, $f_{12}$ and $f_{22}$ are constants and $\mathcal{F}$ is a smooth function that varies slowly in the plane transverse to the direction of motion of the ray. This result is consistent with the temporal decay of the wave amplitude along the ray when $S$ monotonically increases with time.

\section{Modified Standard Cosmological Models}

Starting with the tetrads given in Eqs.~\eqref{H3} and~\eqref{H4}, one can show that $C_{\mu \nu 0} = 0$ and the only nonzero components of the torsion tensor are given by
\begin{equation}\label{B1}
C_{0ij} = - C_{i0j} = \frac{a \dot{a}}{K^2} \delta_{ij}\,, \qquad  C_{ijk} = - C_{jik} = \frac{1}{2}\,\frac{ka^2}{K^3} (\delta_{ik} x^j - \delta_{jk} x^i)\,.
\end{equation} 
It follows that the torsion vector can be expressed as 
\begin{equation}\label{B2}
C_{0}  = -3\,\frac{\dot{a}}{a}\,, \qquad  C_{i} = \frac{k x^i}{K}\,.
\end{equation} 
Moreover, for the contorsion tensor we have $K_{0\mu \nu} = 0$ and the following nonzero components  
\begin{equation}\label{B3}
K_{i0j} = - K_{ij0} = C_{0ij} = \frac{a \dot{a}}{K^2} \delta_{ij} \,, \qquad  K_{ijk} = - K_{ikj} = C_{jki} = \frac{1}{2}\frac{ka^2}{K^3} (\delta_{ij} x^k - \delta_{ik} x^j)\,.
\end{equation} 
Similarly, for the auxiliary torsion tensor we have $\mathfrak{C}_{ij0} = 0$, while the nonzero components are
\begin{equation}\label{B4}
\mathfrak{C}_{0i0} = - \mathfrak{C}_{i00} = C_i\,, \qquad \mathfrak{C}_{0ij} = - \mathfrak{C}_{i0j}  = -2\, C_{0ij}\,, \qquad \mathfrak{C}_{ijk} = - \mathfrak{C}_{jik} = - C_{ijk}\,.
\end{equation} 
Finally, the torsion pseudovector vanishes due to the symmetries of the torsion tensor in this particular case.

With $\check{C} = 0$, the constitutive relation~\eqref{N5} reduces to $N_{\mu \nu \rho} = S\, \mathfrak{C}_{\mu \nu \rho}$; hence, $N_{ij0} = 0$ and the nonzero components of $N_{\mu \nu \rho}$ are given by 
\begin{equation}\label{B5}
N_{0i0} = - N_{i00} = S C_i = \frac{kSx^i}{K}\,, \qquad N_{0ij} = - N_{i0j}  = -2\, S C_{0ij} = - 2\, \frac{Sa \dot{a}}{K^2} \delta_{ij}\,
\end{equation} 
and
\begin{equation}\label{B6}
N_{ijk} = - N_{jik} = - S C_{ijk}  = - \frac{1}{2}\,\frac{kSa^2}{K^3} (\delta_{ik} x^j - \delta_{jk} x^i)\,.
\end{equation} 

Next, we must employ Eqs.~\eqref{T4} and~\eqref{T8} to calculate $Q_{\mu \nu}$ and  $\mathcal{N}_{\mu \nu}$, respectively. 
To compute $Q_{\mu \nu}$ in this case, we first note that 
\begin{equation}\label{B7}
C_{\mu \nu \rho}N^{\mu \nu \rho} = 12 S \left(\frac{\dot{a}}{a}\right)^2 - S\, \frac{k^2 r^2}{a^2}\,.
\end{equation}
We find 
\begin{equation}\label{B8}
Q_{00} = -3\, S \left(\frac{\dot{a}}{a}\right)^2 - \frac{1}{4}\,S\, \frac{k^2 r^2}{a^2}\,,
\end{equation}
\begin{equation}\label{B9}
Q_{0i} = S \,\frac{\dot{a}}{a} \frac{kx^i}{K}\,, \qquad Q_{i0} = 2\,S \,\frac{\dot{a}}{a} \frac{kx^i}{K}\, 
\end{equation}
and 
\begin{equation}\label{B10}
Q_{ij} = - S \left(\frac{\dot{a}}{K}\right)^2 \delta_{ij} - \frac{1}{4}\, \frac{k^2 S x^ix^j}{K^2}\,.
\end{equation}

Similarly, for $\mathcal{N}_{\mu \nu}$, we have
\begin{equation}\label{B11}
\mathcal{N}_{00} = \frac{kS}{a^2}(3K -kr^2)\,,  \qquad \mathcal{N}_{0i} = Q_{0i} + \frac{kx^i}{K} \frac{dS}{dt}\,, \qquad \mathcal{N}_{i0} = Q_{i0}\,
\end{equation}
and 
\begin{equation}\label{B12}
\mathcal{N}_{ij} = - 2\,\frac{1}{K^2}\frac{d(Sa\dot{a})}{d t} \delta_{ij} - \frac{Sk}{K^2}\,\delta_{ij} - \frac{1}{4}\, \frac{k^2 S x^ix^j}{K^2}\,.
\end{equation}

Finally, we can compute the components of $\mathbb{R}_{\mu \nu}$ defined in Eq.~\eqref{T7}. The results are $\mathbb{R}_{0i} = -(x^i/K) k\dot{S}$, $\mathbb{R}_{i0} = 0$ and
\begin{equation}\label{B13}
\mathbb{R}_{00} = -3\, S \left(\frac{\dot{a}}{a}\right)^2 -3\,\frac{Sk}{a^2}\,,
\end{equation}
\begin{equation}\label{B14}
\mathbb{R}_{ij} = -S \left(\frac{\dot{a}}{a}\right)^2\delta_{ij} + 2\,\frac{1}{K^2}\frac{d(Sa\dot{a})}{d t} \delta_{ij} + \frac{Sk}{K^2}\,\delta_{ij}\,.
\end{equation}
Substituting these results in the gravitational field equations, we find that $k \dot{S} = 0$. Therefore, either $k = 0$, as in the Cartesian flat model investigated in Section V, or $S = {\rm constant}$. In the latter case,  the field equations reduce to  Eqs.~\eqref{H5} and~\eqref{H6} of Section VI. 


\section{Modified Cartesian Flat Model Dominated by Dark Energy with $w_{de} < -1$ }

If the model described in Section V is dominated by dark energy with $P = w_{de} \,\rho$, then the dynamical Eqs.~\eqref{F3} and~\eqref{F4} can be combined and expressed as
\begin{equation}\label{C1}
 (1+S) \Big[ 2\frac{\ddot{a}}{a} + \left( \frac{\dot{a}}{a}\right)^2\Big] =  - 3\,w_{de} (1+S) \left( \frac{\dot{a}}{a}\right)^2  - 2 \frac{dS}{dt} \frac{\dot{a}}{a}\,, 
\end{equation} 
where we assume $w_{de} < -1$. To find an explicit solution, the form of $S(t)$ must be specified. 

For the sake of illustration, we assume henceforth that 
\begin{equation}\label{C2}
 S(t) = S_0\, a(t)\,, \qquad w_{de} = -\frac{4}{3}\,,
\end{equation} 
where $S_0 = S(t_0) > 0$ is a positive constant such that $dS/dt > 0$ for an expanding model. It is then straightforward to express Eq.~\eqref{C1} in terms of the deceleration and Hubble parameters given in Eq.~\eqref{f6} as
\begin{equation}\label{C3}
 q  = -\frac{ 3 + S_0\, a}{2(1+S_0\, a)}\,, \qquad \frac{\dot{H}}{H^2} = \frac{1-S_0 \,a}{2(1+S_0 \,a)}\,.
\end{equation} 
The solution for the scale parameter is 
\begin{equation}\label{C4}
 S_0\,a(t)^{1/2} - a(t)^{-1/2} = \frac{t-t_0}{2\,\tau_0} + S_0 -1\,,
\end{equation} 
where $\tau_0$ is a constant of integration with dimensions of time. We are interested in the behavior of this model for $t \to \infty$; therefore, we find that asymptotically
\begin{equation}\label{C5}
a \sim \frac{t^2}{4S_0^2 \tau_0^2}\,, \qquad H \sim \frac{2}{t}\,, \qquad q \sim -\frac{1}{2}\,.
\end{equation} 
Thus, the late-time asymptotic behavior of the modified Cartesian flat model will be an accelerating universe in this case.

\end{document}